\documentclass{ar-1col}
\usepackage[numbers]{natbib}
\usepackage{url}
\usepackage{wrapfig}
\usepackage{float}
\usepackage{bm}
\usepackage{amsmath}
\usepackage{xspace}

\newcommand{\onethree}     {$\sqrt{s}~=~13$~Te\kern-.1emV\xspace}

\newcommand{\forteen}      {$\sqrt{s}~=~14$~Te\kern-.1emV\xspace}

\newcommand{\mt}{\ensuremath{m_{\rm T}}\,} 
\newcommand{\kt}{\ensuremath{k_{\rm T}}\,}

\newcommand{\La}{\ensuremath{\Lambda}\,}

\newcommand{\sig}{\ensuremath{\Sigma}\,}

\newcommand{\xim}{\ensuremath{\Xi^{-}}\xspace}

\newcommand{\omm}{\ensuremath{\Omega^{-}}\xspace}

\newcommand{\kam}          {\ensuremath{\rm{K}^{-}}\xspace}

\newcommand{\pp}{pp\xspace}
\newcommand{\ppNoSpace}{\ensuremath{\mathrm {p\kern-0.05em p}}}
\newcommand{\pPb}{\ensuremath{\mbox{p--Pb}}\,}
\newcommand{\PbPb}{\ensuremath{\mbox{Pb--Pb}}\,}
\newcommand{\AuAu}{\mbox{Au-Au}\xspace}

\newcommand{\LN}{\ensuremath{\mbox{$\Lambda$--N}}\xspace}
\newcommand{\SN}{\ensuremath{\mbox{$\Sigma$--N}}\xspace}

\newcommand{\pP}{\ensuremath{\mbox{p--p}}~}

\newcommand{\pL}{\ensuremath{\mbox{p--$\Lambda$}}~}

\newcommand{\pXim}{\ensuremath{\mbox{p--$\Xi^{-}$}}\xspace}

\newcommand{\pXi}{\ensuremath{\mbox{p--$\Xi$}}\xspace}

\newcommand{\pOm}{\ensuremath{\mbox{p--$\Omega^{-}$}}\xspace}

\newcommand{\NLa}{\ensuremath{\mbox{N--$\Lambda$}}~}

\newcommand{\NXi}{\ensuremath{\mbox{N--$\Xi$}}~}

\newcommand{\NSig}{\ensuremath{\mbox{N--$\Sigma$}}~}
\newcommand{\nSigp}{\ensuremath{\mbox{n--$\Sigma^+$}}~}
\newcommand{\pSigz}{\ensuremath{\mbox{p--$\Sigma^0$}}~}

\newcommand{\Kmp}{\ensuremath{\mathrm {K^-p}}\xspace}
\newcommand{\Kzn}{\ensuremath{\mathrm {\bar K^0n}}\xspace}
\newcommand{\nch}          {\ensuremath{N_{\mathrm{ch}}}\xspace}
\newcommand{\meannch}      {\ensuremath{\langle N_{\mathrm{ch}}\rangle}\xspace}
\newcommand{\ks}           {\ensuremath{k^*}\xspace}
\newcommand{\rs}           {\ensuremath{r^*}\xspace}
\newcommand{\sr}         {\ensuremath{S(r^*)}\xspace}

\newcommand{\CF}           {C(\ks)}

\newcommand{\MeVc}{\ensuremath{\mathrm{MeV}\kern-0.05em/\kern-0.02em \textit{c}}~}
\newcommand{\GeVc}{\ensuremath{\mathrm{GeV}\kern-0.05em/\kern-0.02em \textit{c}}~}
\newcommand{\GeVcSq}{\ensuremath{\mathrm{GeV}\kern-0.05em/\kern-0.02em \textit{c}^2}~}
\newcommand{\MeVcSq}{\ensuremath{\mathrm{MeV}\kern-0.05em/\kern-0.02em \textit{c}^2}~}

\newcommand{\radiuspOmega}{\ensuremath{0.95\pm0.06}\,}

\newcommand{\chiEFT}       {\ensuremath{\chi}\rm{EFT}\xspace}

\jname{Annu.Rev.Nucl.Part.Sci.}
\jvol{77}
\jyear{2021}
\doi{10.1146/annurev-nucl-102419-034438}

\begin{document}

\markboth{Author et al.}{Short title}

\title{Study of the Strong Interaction Among Hadrons with Correlations at the LHC}

\author{L. Fabbietti,$^1$ V. Mantovani Sarti,$^1$ and O. V\'azquez Doce$^1$
\affil{$^1$ Physik Department Technische Universit{\"a}t M{\"u}nchen, 85478 Garching, Germany; email: laura.fabbietti@ph.tum.de}}

\begin{abstract}
The strong interaction among hadrons has been measured in the past by scattering experiments. Although this technique has been extremely successful in providing information about the nucleon–nucleon and pion–nucleon interactions, when unstable hadrons are considered the experiments become more challenging. In the last few years, the analysis of correlations in the momentum space for pairs of stable and unstable hadrons measured in pp and p-Pb collisions by the ALICE Collaboration at the LHC has provided a new method to investigate the strong interaction among hadrons. In this article, we review the numerous results recently achieved for hyperon–nucleon, hyperon–hyperon, and kaon–nucleon pairs, which show that this new method opens the possibility of measuring the residual strong interaction of any hadron pair.
\end{abstract}

\maketitle

\tableofcontents

\section{Introduction}
The study of the residual strong interaction between hadrons (colorless bound states of quarks and anti-quarks) is still an open topic in nuclear physics. While the Standard Model of elementary particle physics provides a satisfactory description at the quark level in the high-energy regime, the low energy processes that characterize the interaction among hadronic degrees of freedom, are not yet described by a fundamental theory and are often difficult to access experimentally. 

Hadron--hadron interactions have been studied in the past by means of scattering experiments at low energies (below the nucleon mass) for both stable and unstable beams. A reasonable amount of scattering data (roughly 8000 data points)~\cite{Arndt:2007qn,Perez:2013mwa} are available for nucleon--nucleon (NN) reactions, but for kaons (mesons with one strange quark) and hyperons (baryons with at least one  strange quark) the beam realization is more challenging. Kaon--nucleon interactions can be studied because the necessary secondary beams are accessible \cite{Mast:1975pv,Ciborowski:1982et}, but for hyperon--nucleon reactions the data is scarcer (17 data points) \cite{Eisele:1971mk,Alexander:1969cx,SechiZorn:1969hk} because of the unstable nature of the hyperon beams due to weak and electromagnetic decays.
The lack of statistics in reactions involving unstable hadrons affects the current description of the corresponding strong interaction from a theoretical point of view. If we consider only hadrons containing u,d and s quarks, most of the predicted interactions are not constrained experimentally. This represents not only a limit for nuclear physics but has also implications for astrophysics. Neutron stars (NS), for example, could be constituted from nucleons, hyperons and kaons and their properties strictly depend upon the interactions among these hadrons \cite{Weissenborn:2011kb,Lonardoni:2014bwa,Gerstung:2020ktv}. Although the dense environment present within neutron stars is not easy to realize under controlled conditions with terrestrial experiments, the study of two- and three-body interactions among neutrons, protons, hyperons and kaons in vacuum drive the equation of state (EoS) of NS.

In this review we focus on the novel input provided by the ALICE experiment to the topic of the interactions among nucleons, hyperons and kaons by means of the femtoscopy method applied to data from ultra-relativistic pp and p--Pb collisions at the LHC.\\

Historically, the femtoscopy technique can be traced back to the first measurements of particle interferometry with photons, performed by Hanbury, Brown and Twiss during the 1950s, used to determine the size of stars~\cite{HanburyBrown:1956bqd}. The same idea has been subsequently applied to pairs of identical particles in elementary and heavy-ion collisions (HIC)~\cite{Goldhaberpions,Gyulassypions,Bevalac1,Bevalac2,Wiedemann:1999qn,Podgoretsky:1989bp}, proving to be an extremely useful tool to determine the space-time structure of the emitting source. The analysis of pion or kaon pairs, where the quantum statistics together with the Coulomb interaction characterize the shape of the correlations, has dominated the femtoscopy studies in the last three decades.
 Results from intermediate-energy heavy-ion collisions (HIC) at the Bevalac in the mid 1980s quantitatively showed that the spatial dynamics of the system could be probed by femtoscopy and in the following years femtoscopic studies have been performed in several different experiments and energy ranges, from SIS ($\sqrt{s_{\mathrm{NN}}}=$ 1-3 GeV) ~\cite{Hades1,Hades2}, AGS ($\sqrt{s_{\mathrm{NN}}}=$ 5-10 GeV)~\cite{AGS1,AGS2}, SPS ($\sqrt{s_{\mathrm{NN}}}=$ 17 GeV)~\cite{SPS1,SPS2}, and more recently RHIC ($\sqrt{s_{\mathrm{NN}}}=$ 200 GeV) \cite{Heinz:2002un} to LHC ($\sqrt{s_{\mathrm{NN}}}=$ 5-13 TeV)\cite{Khachatryan:2010un}.\\
The abundance of collected data made it possible to study the three-dimensional evolution of the particle emitting source, helping to characterize the kinematic freeze out of different hadron species. 
The typical source sizes measured in HIC ranges from $2$ fm for SIS energies up to $5-6$ fm  for measurements at RHIC and at the LHC \cite{Chojnacki:2011hb,PhysRevC.92.054908}.\\
In the last two decades, great effort has been put into using femtoscopy to understand and study the strong interaction among hadrons~\cite{Lednicky:2003mq}.
This line of research was pioneered in the last decade \cite{Shapoval:2014yha,Hades1}
and then developed further by the STAR collaboration studying $\Lambda$--$\Lambda$~\cite{LambdaLambdaSTAR}, $\bar{\mathrm{p}}$--$\bar{\mathrm{p}}$~\cite{apapSTAR} and lately p--$\Omega^-$~\cite{STAR:2018uho} correlations measured in Au--Au collisions at $\sqrt{s_{NN}}=200$ GeV. These studies showed the limits of the method applied to ultrarelativistic heavy-ion collisions where the average inter-particle distances of 7-8 fm reduces the sensitivity to the short-range strong interaction. This distance of particles at kinetic freeze out is obtained from the probability density distribution for Gaussian sources with radii of 3-4 fm \cite{Lisa:2005dd}. They also demonstrated that the interaction studies require an extremely high purity for particle identification and a detailed treatment of the residual background \cite{Morita:2014kza}. Following the same approach, the ALICE collaboration successfully extracted for the first time the $\Lambda$--K and baryon--antibaryon scattering parameters from \PbPb collisions at $\sqrt{s_{NN}}=2.76$ and $5.02$ TeV~\cite{LambdaKaonALICE,BBAr_HICALICE}. The results by the ALICE and STAR collaborations provided a first proof that the correlation function can be exploited to infer information on the underlying strong interaction.\\
In the last three years, the ALICE collaboration applied the femtoscopy technique also to pp and p-Pb collisions and showed for the first time the potential to precisely assess the strong interaction amongst stable and unstable hadrons. It was demonstrated that the average inter-particle distance obtained from such collisions at the LHC is about 1 fm and hence comparable to the range of the strong potentials. This feature, combined with the excellent particle identification provided by the ALICE apparatus and the extensive statistics collected during the LHC Run 2 data taking, allowed to precisely measure the following interactions: p--p~\cite{FemtoRun1}, K$^+$--p and K$^-$--p~\cite{Acharya:2019bsa}, p--$\Lambda$~\cite{FemtoRun1}, p--$\Sigma^0 $~\cite{Acharya:2019kqn}, $\Lambda$--$\Lambda$\cite{FemtoLambdaLambda}, p--$\Xi^-$~\cite{FemtopXi}, and p--$\Omega^-$~\cite{Acharya:2020asf}.
This review describes the main features of the femtoscopy technique, the advantages of using it in small colliding systems and the results obtained in the study of hadron--hadron interactions with strangeness.

The review is structured as follows: the femtoscopy method is presented in Sec.~\ref{sec:Methodology}. In Sec.~\ref{sec:source}, the modeling of the emitting source in pp collisions is presented and the main features of femtoscopy in elementary collisions are discussed.
The main results on hadron-hadron interactions obtained with the ALICE femtoscopy measurements are discussed in Sec.~\ref{sec:stronginteraction} with particular emphasis on hyperon-nucleon systems (Sec.~\ref{subsec:hypnuclint}), the possible detection of bound states (Sec.~\ref{subsec:boundstates}) and coupled-channel dynamics (Sec.~\ref{subsec:coupledchannel}). In Sec.~\ref{sec:neutronstars} the possible implications for the presence of hyperons inside neutron stars are  
considered. Finally in Sec.~\ref{sec:outlook} the future prospects for femtoscopy achievements in the next ALICE data-taking
periods (Run 3, Run 4) are discussed and Sec.~\ref{sec:summary} summarizes the current state of the field and points to future theoretical and experimental developments.

\section{Methodology}
\label{sec:Methodology}
The fundamental quantity to be measured in femtoscopy is the correlation function. It is expressed as a function of the relative distance between two particles $\rm \mathbf{r^*}$ and their reduced relative momentum, $\ks\,=\,|\rm \mathbf{p^*_2}-\rm \mathbf{p^*_1}|$/2~ in the pair rest frame, with $\rm \mathbf{p^*_1} = - \rm \mathbf{p^*_2}$, by the Koonin-Pratt formula~\cite{Pratt:1986cc,Lisa:2005dd}\footnote{The relative momentum formula in terms of the invariant relative momentum $q_{inv}$ reads
$k^* = \sqrt{\frac{a^2-m_1^2m_2^2}{2a+m_1^2+m_2^2}}$,
$a= \frac{1}{2}(q_{inv}^2 +m_1^2 + m_2^2)$ 
and 
$q_{inv}^2= \mid \rm \mathbf{p}_1 - \rm \mathbf{p}_2 \mid^2 - \mid E_1-E_2 \mid^2$, with $m_1$ and $m_2$ the mass of the particles in the pair and $\rm \mathbf{p}_{1,2}$ the particle momenta in the laboratory reference system. If the two particles have the same mass then $k^* = \frac{q_{inv}}{2}$.}
\begin{equation}
    C(k^*) = \int S(\rm \mathbf{r^*}) |\psi(\rm \mathbf{r^*},\rm \mathbf{k^*})|^2 d^3r.
\label{eq:KooninPratt}
\end{equation}
The first term in Eq. \ref{eq:KooninPratt}, $S(\rm \mathbf{r^*})$, describes the source emitting particles; the second term contains the interaction part via the two-particle wave function $\psi(\rm \mathbf{r^*},\rm \mathbf{k^*})$.
The shape of the correlation function will be determined by the characteristics of the source function and the sign and strength of the interaction. \\
An analytical model by Lednický and Lyuboshitz~\cite{Lednicky:1981su} exists to compute this correlation function.
The Lednický-Lyuboshitz (LL) model assumes a Gaussian profile that depends only on the magnitude of the relative distance for the source function
\begin{equation}
    S\left(r^{*}\right)=\left(4 \pi r_{0}^{2}\right)^{-3 / 2}\cdot \exp \left(-\frac{{r^*}^{2}}{4 r_{0}^{2}}\right),
\label{eq:lednicky_source}
\end{equation}
where $r_0$ is the radius parameter that defines the size of the source.
The effective range approximation is used to define the complex scattering amplitude as

\begin{equation}
    f\left(k^{*}\right)^{S}=\left(\frac{1}{f_{0}^{S}}+\frac{1}{2} d_{0}^{S} k^{* 2}-i k^{*}\right)^{-1},
    \label{eq:lednicky_scatteringamplitude}
\end{equation}
with $S$ the total spin of the particle pair, and $f_0^S$ and $d_0^S$ the scattering length and the effective range, respectively.
The correlation function for uncharged particles becomes then 

\begin{multline}
    C\left(k^{*}\right)_{\text {LL}}= 1+\sum_{S} \rho_{S}\Bigg[\frac{1}{2}\left|\frac{f\left(k^{*}\right)^{S}}{r_{0}}\right|^{2}\left(1  -\frac{d_{0}^{S}}{2 \sqrt{\pi} r_{0}}\right)+\nonumber \\
+ \frac{2 \Re f\left(k^{*}\right)^{S}}{\sqrt{\pi} r_{0}} F_{1}\left(2\ks r_{0}\right) 
    -\frac{\Im f\left(k^{*}\right)^{S}}{r_{0}} F_{2}\left(2\ks r_{0}\right)\Bigg],
\end{multline}    
\label{eq:lednicky_cf}
where $F_1(2\ks r_0)$ and $F_2(2\ks r_0)$ are analytical functions which result from the Gaussian source approximation,
and $\rho_S$ is the pair fraction that is emitted into the spin state S.\\
Since the LL approach is based on the effective range expansion it presents limitations for small systems, because it does not account for the details of the wave function at small distances where the effect of the strong potential is more pronounced. \\
Such limitations motivated the development of the \textbf{C}orrelation \textbf{A}nalysis \textbf{T}ool using the \textbf{S}chrödinger equation" (CATS) framework~\cite{Mihaylov:2018rva}, which provides a numerical recipe for the calculation of the exact solution of the two-body non--relativistic \textbf{S}chrödinger equation given a local interaction potential. The resulting relative wave function combined with a parametrization of the source makes it possible to compute predictions for the different correlation functions by means of Eq.~\ref{eq:KooninPratt}. The CATS framework is able to account for both short range potentials and the Coulomb long range interaction, as well as different parametrizations of the source function beyond the Gaussian approximation. Moreover the LL model is also implemented in CATS, using the the scattering length $f_0^S$ and the effective range $d_0^S$ as input for the description of the interaction.

The features of the interaction are mapped into the corresponding correlation function. In particular, the effects of the final-state interactions are more evident in the correlation function at small \ks values. A repulsive interaction, with positive values of the local potentials will imply a correlation function with values between 0 and 1. For an attractive interaction instead, the resulting correlation function gets values above unity.
This intuitive picture is modified, though, if the attraction is strong enough to accommodate the presence of a bound state. In this case a depletion in the values of the correlation function can be seen, depending on the binding energy (BE)~\cite{Morita:2019rph}. This depletion is due to the fact that the pairs that form the bound state are lost to the correlation, since they result into a different final state.

\begin{figure}[h!]
  \centering \includegraphics[width=\textwidth]{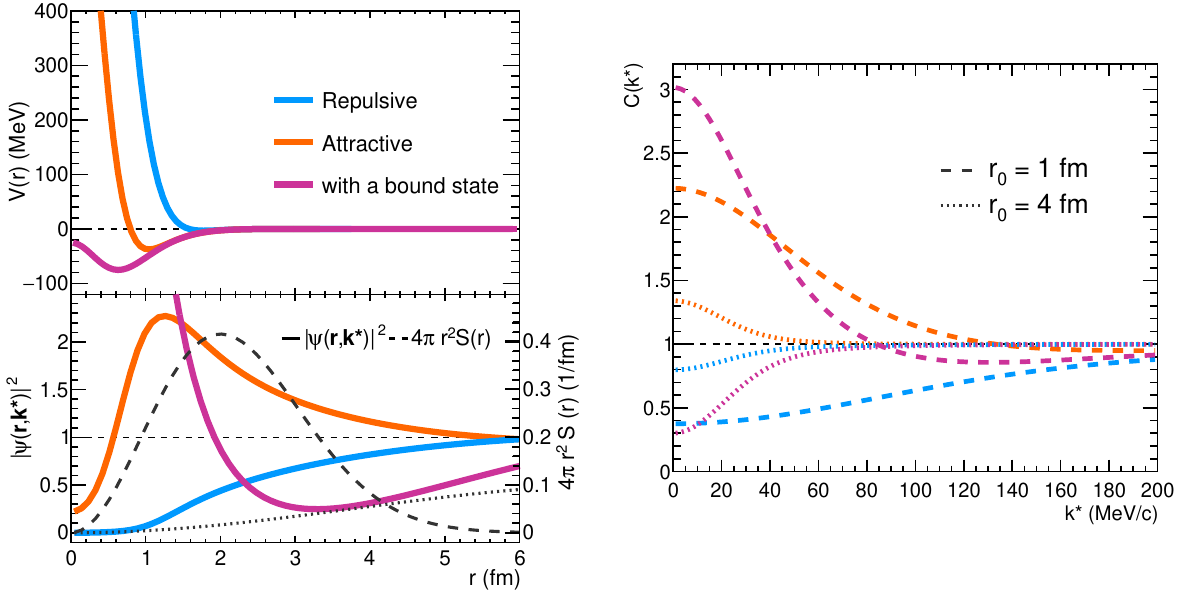} \caption{\textit{(Color online)}. Left upper panel: Examples of an attractive (orange), repulsive (azure) and a potential with a shallow bound state (pink). Left lower panel: Modulus squared of the total wave-functions $|\psi(\rm \mathbf{r^*},\rm \mathbf{k^*})|^2$, as a function of the relative distance r, obtained by solving the Schr\"odinger equation employing the CATS framework for the three example potentials. On the same plot the profile of the emitting source is shown for $1$ fm (dashed black) and $4$ fm (dotted black). Right panel: resulting correlation function C(\ks) for each interaction, evaluated for the two different source sizes $r_0 = 1, 4$ fm.}
\label{fig:Examples}
\end{figure}

The strength of the correlation can also be enhanced by small sizes of the source function, as  will be discussed.
Other effects, not caused by the final-state interaction, can be visible at different \ks ranges of the correlation function, like quantum-mechanical interference, resonances, or conservation laws.\\
Figure \ref{fig:Examples} demonstrates the sensitivity of the 
femtoscopy method applied to small colliding systems for the study of the strong interaction assuming an attractive, repulsive or binding potential.
The left lower panel shows the corresponding  squared modulus of the wavefunction
together with the density distribution according to Gaussian profiles with two different radii ($r_0= $ 1 and 4 fm). The sensitivity of the method to study the strong interaction depends on the overlap of the square of the wave function with the source density distribution. Typical sizes of the parameter $r_0= $ 3--6 fm describe the source formed in Pb-Pb collisions at the LHC~\cite{Acharya:2017qtq}, while pp and p-Pb collisions at the same energies lead to the formation of sources with much smaller radii between 1--1.5 fm~\cite{FemtoRun1,FemtopXi}. One can see that for heavy-ion collisions the overlap is minimal and hence the sensitivity to the short range interaction is very limited.
The typical features of the attractive and repulsive interactions and the presence of the bound state are much more pronounced in the case of the small source size. The less pronounced correlation function obtained with the larger source is very difficult to measure with sufficient precision.\\
In the case of the bound state, the reason why the correlation function flips around 1 for different source sizes is due to the fact that the wave function is very sharply peaked towards distances equal to zero, due to the much stronger localization of the bound state. This translates into an increased correlation for small radii, while for large radii only the asymptotic part of the wave function, which is depleted because of conservation of probability, impacts the correlation function and bring it below one.

The method used by ALICE to study the interactions among hadrons consists in the comparison of the theoretical expectation for the correlation function to a correlation function obtained experimentally.
The experimental correlation function is
obtained as the ratio of the relative momentum distribution of pairs of particles produced in the same event (SE), that constitutes the sample of correlated pairs, with a reference distribution obtained combining particles produced in different collisions using the so-called mixed event (ME) technique
\begin{equation}
    C(\ks)\,=\,\xi(\ks)\cdot\frac{N_{\mathrm{SE}}(\ks)}{N_{\mathrm{ME}}(\ks)}.
    \label{eq:cf_experimental}
\end{equation}
The corrections for experimental effects are denoted by $\xi(\ks)$ in Eq. \ref{eq:cf_experimental}. Such corrections take into account the finite experimental resolution, and corrections to the ME distributions in order to assure the same experimental conditions and normalization as for the SE events.  In general, they do not account for the contributions from misidentification, weak decays or residual background induced by mini-jets and event-by-event momentum conservation. These effects are accounted for in the fit of the correlation functions.

The experimental correlation function is further distorted by two distinct mechanisms. The sample of particle pairs can include, beside primary particles, also misidentified particles and secondary particles from weak decays of resonances. This introduces contributions of different, non-genuine, pairs in the measured correlation function. The treatment of these contributions is described in detail in~\cite{FemtoRun1}, and therefore here we only briefly sketch the procedure. The contributions of the different non-genuine and genuine correlation to the total experimental correlation are indicated by weights called $\lambda$ parameters.
These $\lambda$ parameters are obtained by pairing single particle properties such as the purity (P) and feed-down fractions (f): $\lambda_{ij}=P_i P_j f_i f_j$.
The total correlation function can then be decomposed as: 
\begin{equation}\label{eq:lambdapars}
C(k^*) = 1 + \lambda_{\mathrm{genuine}} \cdot [ C_{\mathrm{genuine}}(k^*)-1 )
+\sum_{ij}\lambda_{ij}(C_{ij}(k^*) -1],
\end{equation}
where the $i,j$ denote all possible impurity and feed-down contributions.
\begin{marginnote}
\entry{Primary and secondary particles}{The fraction $f_i$ of primary and secondary (i.e. produced in a decay) particles  entering the $\lambda_{ij}$ parameters are evaluated from experimental data considering weak, electromagnetic decays and fake candidates for each hadron species.}
\end{marginnote}

The correlation functions measured by ALICE are compared with theoretical expectations obtained according to Eq. \ref{eq:lambdapars}.
For this task the $\lambda$ parameters are obtained from experimental data when possible (e.g.: purity, fractions of secondary particles). In addition, the experimental effects denoted by $\xi(\ks)$ in Eq. \ref{eq:cf_experimental} are taken into account when modelling the theoretical correlation function.
The only exception is the study of the \pXi and \pOm correlation functions in \pp collisions at 13 TeV published in~\cite{Acharya:2020asf}, where the experimental data have been unfolded for all effects and it is directly compared with the genuine theoretical correlation function.\\
In order to account for residual contributions by mini-jets background to the final correlations, a baseline with free parameter is multiplied to the correlation function $C(k^*)$ used to fit the experimental data \cite{FemtoRun1}. This baseline assumes different shapes depending on the pair of interest, but contributes at most few percent to the global correlation strength \cite{FemtopXi,Acharya:2019kqn}.

\section{Determination of the particle emitting source}
\label{sec:source}
After the collision and after the hadronization processes are completed, particles might undergo some inelastic collisions but shortly after their production they propagate freely towards the detectors. The distribution of the space coordinates at which the different particles assume their primary momentum values characterize the particle emitting source.
The understanding of this source for the selected colliding system is mandatory in order to extract informations on the underlying strong interaction.

Gaussian source profiles in one and three dimensions are typically assumed in femtoscopic studies performed in heavy-ion collisions~\cite{Lisa:2005dd, Lisa:2008gf}. However, the presence of a collective expansion can introduce correlations between the position and the momentum of the emitted particles. This effect can be seen as a decrease of the extracted gaussian radii with increasing pair transverse momentum \kt~\cite{Lisa:2005dd,Bearden:2000ex}. Experimentally,
a common scaling of the source size with the transverse mass of protons and kaons pairs has been seen in \PbPb collisions~\cite{Adam:2015vja}.

High-multiplicity pp collisions and heavy-ion systems exhibit similar behavior in several related measured quantities, such as angular correlations and strangeness production~\cite{Khachatryan:2016txc,Khachatryan:2010gv,ALICE:2017jyt,Acharya:2018orn}. Hence, a similar transverse mass \mt scaling\begin{marginnote}
\entry{Transverse mass}{The transverse mass of the pair is defined as $\mt = \sqrt{\kt^{2} + m^2}$, where $m$ is the average mass of the particle pair and $\kt = \frac{\mid \rm \mathbf{p}_{\rm{T},~1} + \rm \mathbf{p}_{\rm{T},~2}\mid}{2}$ is the relative transverse momentum.}
\end{marginnote}of the source size as observed in large systems is expected to apply in pp collisions. Measurements of this kind in small systems are currently available for light meson pairs ($\pi$--$\pi$, K--K)~\cite{Sirunyan:2017ies,Aad:2015sja,Abelev:2012sq,Abelev:2012ms,Sirunyan:2019umv}, accessing only low values of \mt, but indicating already a dependence of the radius on the transverse mass.

Recently, similar studies of baryon-baryon femtoscopy, for p--p and p--$\Lambda$ pairs, have been conducted in high-multiplicity pp collisions and these studies provided for the first time a quantitative measurement of a common scaling in the range $\mt = 1.3 - 2.4$ \GeVcSq~\cite{Acharya:2020dfb}. The explicit inclusion of strongly decaying resonances proved to be a fundamental ingredient for the description of the data.
The presence of feed-down from strong resonances had already been suggested as a possible explanation for the description of the different scaling of radii extracted in $\pi\mbox{--}\pi$ correlations seen in heavy-ion collisions~\cite{Sinyukov:2015kga,Wiedemann:1996ig}. A similar broken scaling is observed in pp collisions for \pL pairs when a Gaussian source profile is assumed and feed-down effects are not taken into account~\cite{Acharya:2020dfb}. The extracted \pL radii are typically $20\%$ larger with respect to the \pP pair results.
In the spirit of testing the hypothesis of a common source for small colliding systems, a complete modeling of the resonance contributions has to be performed.

The emitting source \sr used to fit \pP and \pL correlation functions with Eq.~\ref{eq:KooninPratt} is composed of a Gaussian core of width $r_\mathrm{core}$ (see Eq.~\ref{eq:lednicky_source}), related to the emission of all primordial particles


\begin{align}
    S_{\rm prim}\left(\rm \mathbf{r}^{ \,*} _{\rm core}\right)=\left(4 \pi r_{\rm core}^{2}\right)^{-3 / 2}\cdot \exp \left(-\frac{\rm \mathbf{r}^{\,*\,\,\,\,2}  _{\rm core}}{4 r_{\rm core}^{2}}\right) \,\,\,\,, 
\end{align}

and of an exponential distribution originating from strong decays of resonances with a specific lifetime $\tau_{\rm res}$.\\
The modification of the relative distance $\rm \mathbf{r}^{\,*}$ of the particle in the pair, entering the final description of the source, linearly depends on both the core distance $\rm \mathbf{r}^{\,*} _{\rm core}$ and on the distances $\rm \mathbf{r}^{\,*} _{\rm res,i}$ traveled by the resonances $i=1,2$ of momentum $\rm \mathbf{p}^{\,*}_{\rm res,i}$, mass $M_{\rm res,i}$ and flight time $t_{\rm res,i}$ sampled from the exponential distribution based on the corresponding lifetime $\tau_{\rm res,i}$:

\begin{align}\label{eq:vectorsSource}
    \rm \mathbf{r}^{\,*} = \rm \mathbf{r}^{ \,*} _{\rm core} + \sum_i \rm \mathbf{r}^{ \,*} _{\rm res,i} \,\,\, ,\,\,\,
    \rm \mathbf{r}^{\,*} _{\rm res,i} = \frac{\rm \mathbf{p}^{\,*}_{\rm res,i}}{M_{\rm res,i}}t_{\rm res,i}.
\end{align}

These latter quantities, related to the resonances feeding to p and \La, depend on the resonance yields and kinematics.
The absolute value $r^*=|\rm \mathbf{r}^{\,*}|$ needs to be evaluated for the one-dimensional source function $S(r^*)$ . From the definitions in Eq.~\ref{eq:vectorsSource}, the needed ingredients are $r^*_\mathrm{core}$, the momenta, masses and lifetimes of the resonances, as well as the angles formed by  $\rm \mathbf{r}^{\,*}_\mathrm{core}$, and the resonance distances $\rm \mathbf{s}^{\,*}_\mathrm{res,1}$ and $\rm \mathbf{s}^{\,*}_\mathrm{res,2}$.

\begin{figure}[htb]
 \includegraphics[width=\textwidth]{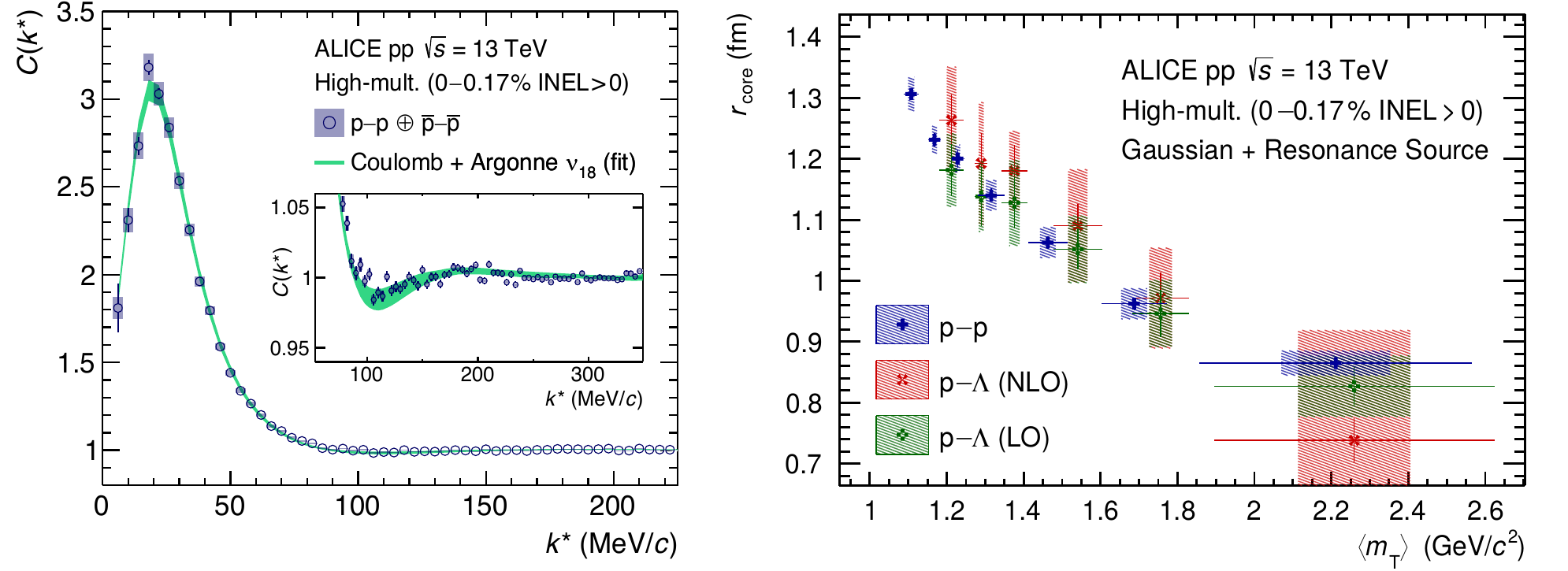}
  \caption{\textit{(Color online)}. In both plots: statistical (bars) and systematic uncertainties (boxes) are shown separately. Left: \mt integrated \pP correlation function as a function of \ks measured in high-multiplicity pp collisions, including the contributions from strong resonances. The width of the band (green) represents one standard standard deviation of the systematic uncertainty of the fit. Right: Gaussian core radius $r_{\textrm{core}}$ as a function of $\langle\mt\rangle$. Blue crosses correspond to the \pP correlation function fitted with Argonne $v_{18}$ potential~\cite{Wiringa:1994wb} as the strong potential. The green squared crosses (red diagonal crosses) result from fitting the \pL correlation functions with the strong $\chi$EFT LO~\cite{Polinder:2006zh} (NLO~\cite{Haidenbauer:2019boi})}
  \label{fig:ppHM13_Source}
\end{figure}

The amount and type of resonances can be estimated from calculations based on the statistical hadronization model~\cite{Vovchenko:2019pjl,Becattini:2001fg,Wheaton:2004qb} and are found to be similar ($\approx 65\%$) between the two particles, coming mostly from $\Delta$ for protons, and from $\Sigma^*$ for \La. This different composition of secondary particles translates into a significantly larger average lifetime and average mass ($M_{res} =1.46$ \GeVcSq, $c\tau_{res} = 4.7$ fm) for \La hyperons with respect to protons ($M_{res} =1.36$ \GeVcSq, $c\tau_{res} = 1.6$ fm), explaining qualitatively the larger effective source obtained in~\cite{Acharya:2020dfb}.
The remaining kinematics parameters, namely the momenta of the resonances and their relative orientation with respect to $\rm \mathbf{r}^{\,*}_\mathrm{core}$, are determined from transport model simulations within the EPOS framework~\cite{Pierog:2013ria}.

The total source can be finally decomposed, depending on the origin of each particle in the pair (either primary or from a resonance), as follows:

\begin{align}
 \sr = & P_{\rm prim} P_{\rm prim} \times S_{\rm prim-prim}(\rs)+
 P_{\rm prim} P_{\rm res} \times S_{\rm prim-res}(\rs)+\nonumber\\
 &P_{\rm res} P_{\rm prim} \times S_{\rm res-prim}(\rs)+
P_{\rm res} P_{\rm res} \times S_{\rm res-res}(\rs).
\end{align}\label{eq:DecompSource_CoreRes}

Here $P_{\rm prim}$ and $P_{\rm res}$ are the fractions of primordial and resonance contributions estimated from thermal model calculations. 
Once all the resonance dynamics and composition for the considered pair are accounted for, the Gaussian source size $r_{\rm core}$, related to the prompt emission of particles, remains as the only free parameter to be determined via a fit to the data.

A differential \mt analysis has been performed on \pP and \pL correlations and the core radius has been extracted in each \mt bin.
In the left plot of Fig.~\ref{fig:ppHM13_Source}, the resulting \mt integrated \pP correlation function is shown, obtained assuming the core-resonance source model. The genuine \pP term of the correlation is modeled using the CATS framework, assuming the Argonne $v_{18}$~\cite{Wiringa:1994wb} as the strong potential (including S,P and D waves), and including the Coulomb interaction along with the proper quantum statistics anti-symmetrization of the wave-function. The underlying strong interaction between protons is known with high precision and it is accurately described by the Argonne $v_{18}$ potential~\cite{Wiringa:1994wb}, allowing for a reliable determination of the $r_{\rm core}$ parameter. The data are nicely reproduced by the modeled correlation and the same fitting procedure has been adopted in the single \mt bins, leading to similar results. The \pL interaction is less constrained~\cite{FemtoRun1,Hashimoto:2006aw,SechiZorn:1969hk,Alexander:1969cx,Eisele:1971mk}, hence both leading order (LO)~\cite{Polinder:2006zh} and next-to-leading order (NLO)~\cite{Haidenbauer:2019boi} chiral effective field theory calculations ($\chi$EFT) have been employed.\\
In the right plot of Fig.~\ref{fig:ppHM13_Source} the \mt dependence of the extracted core radii for the two pairs is presented. As can be clearly seen, the inclusion of resonances in the modeling of the source provides a common \mt scaling for both baryon-baryon pairs, providing the first quantitative evidence of a common emitting source in small systems.
This results represents a fundamental input in the investigation of the strong interaction by means of femtoscopy, since it allows to fix the source for any baryon-baryon pair, given the $\left <\mt\right>$ of the pair and the resonance contributions.

If the source $S(r^*)$ is under control from the analysis of particle species for which the 
final state interaction is known, then the relative wave function $\psi(\rm \mathbf{r^*},\rm \mathbf{k^*})$ and hence the interaction for other species can be determined by studying the correlation function (see Eq. 1). Also, the small values found for the core radius $r_{\rm core}$ implies that pp collisions at the LHC are excellent systems to study to short range strong interaction.



\section{Probing the strong interaction for strange hadrons}
\label{sec:stronginteraction}
The interaction between strange hadrons and nucleons is not very well constrained by experimental data. 
In particular, the unstable nature of hyperons makes it very complicated to investigate two- and three-body interactions. 
The high statistics that have been collected for all hyperon species in pp and p-Pb collisions measured by the ALICE collaboration during Run 1 and Run 2 at the LHC allowed unprecedented precision in the study of different interactions, including combinations that had been studied already in past by means of scattering experiments (p--K$^{+,-}$ and p--$\Lambda$)~\cite{Humphrey:1962zz, Watson:1963zz, Nowak:1978au,Hadjimichef:2002xe,SechiZorn:1969hk,Alexander:1969cx,Eisele:1971mk,Hashimoto:2006aw} or were never measured before (p--$\Xi^-$ and p--$\Omega^-$).

The identification, tracking and momentum resolution provided by ALICE for all charged particles makes it possible to study correlation functions down to relative momenta of $4-10$ MeV/$c$. A precise measurement of the correlation function in this momentum range is necessary to study the details of the strong interaction. Moreover the large amount of hyperons, including species such as $\Xi$ and $\Omega$, makes it possible to measure hadron pairs not accessible in the standard scattering experiments. The measurement of these hyperons also allow to test predictions from lattice QCD for interactions with nucleon, since for such heavy hyperons the calculation results are rather solid \cite{Sasaki:2019qnh}.
\\
The femtoscopic measurements performed in small colliding systems such as pp and p-Pb grant access to the short-range strong interaction, as already shown in Sec.~\ref{sec:Methodology}. In the following sections we will discuss in detail different features such as coupled-channel effects and formation of bound states, arising from the short-range dynamics of hadron-hadron potentials.  

\subsection{Study of the hyperon-nucleon interaction}\label{subsec:hypnuclint}
\begin{figure}[h]
\centering
\includegraphics[width=0.999\textwidth]{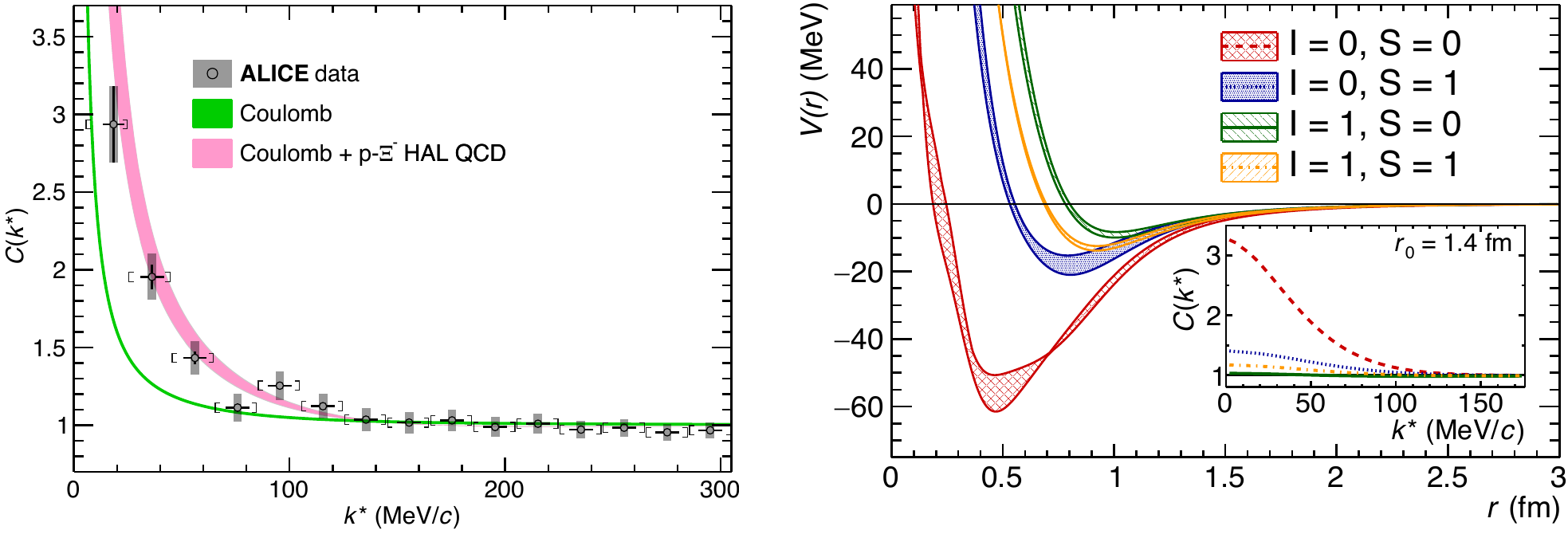}
  \caption{\textit{(Color online)} Left panel: p--$\Xi^-$ correlation function measured in pp collisions at $\sqrt{s_{NN}}= 13$ TeV~\cite{Acharya:2020asf} recorded with a high multiplicity trigger. The experimental data are shown by the black symbols together with statistical and systematic errors. The green curve represents the predicted correlation function assuming only the Coulomb interaction. The pink curve shows the prediction obtained considering the Coulomb and strong interaction provided by the HAL QCD group~\cite{Sasaki:2019qnh}. Right panel: strong potentials for the different spin and isospin configuration of the p--$\Xi^-$ interaction as a function of the inter-particle distance~\cite{FemtopXi}. }
  \label{fig:pXi}
\end{figure}

One of the challenging measurements achieved applying the femtoscopy technique to pp and p-Pb collisions and interpreting the observables with the help of the CATS framework is the study of the p--$\Xi^-$ interaction. 
The $\Xi^{\pm}$ hyperons are reconstructed exploiting the weak decays  $\Xi^{\pm}\rightarrow \Lambda + \pi^{\pm}$ and $\Lambda \rightarrow \mathrm{p} +\pi^-$. A total invariant mass resolution below $2$ MeV/$c^2$~\cite{Acharya:2019sms} is obtained for the reconstructed $\Xi^{\pm}$ and the obtained p--$\Xi^- \oplus\bar{\mbox{p}}$--$\Xi^+$
correlation function is shown in the left panel of Fig.~\ref{fig:pXi}. The data have been corrected for experimental effects.
The right panel of Fig.~\ref{fig:pXi} shows the strong potentials predicted by the HAL QCD collaboration for the four allowed spin and isospin states of the p--$\Xi^-$ system. One can see that for all cases an attractive interaction and a repulsive core characterize the potentials. 

All the potentials are similar at inter-particle distances above $1.5$ fm but the inset shows that the corresponding correlation functions, that have been evaluated using a Gaussian source with a radius equal to $r_0=\, 1.4$ fm~\cite{FemtopXi}, are very different.
This difference is due to the sensitivity of the method to small distances, below 1 fm, that are typical for pp and p-Pb collisions.
The total \pXim correlation function shown in the left panel of Fig. \ref{fig:pXi} is obtained including the strong and the Coulomb potentials in the Schr\"odinger equation and combining the correlation functions for each of the allowed spin and isospin states weighted by the proper Clebsch-Gordon coefficients, following\\
  $C_{\pXim} =  \frac{1}{8}C_{(I=0,\,S=0)} + \frac{3}{8}C_{(I=0,\,S=1)} +  \frac{1}{8}C_{(I=1,\,S=0)} + \frac{3}{8}C_{(I=1,\,S=1)}.$
The source size for the \pXim pair has been evaluated following the model described in Sec.~\ref{sec:source}, considering the average $m_{\mathrm{T}}$ of the pair and the strong resonance contribution for the protons, leading to a value of $r_0 = \, 1.02 \pm 0.05$ fm. 
The total \pXim correlation for the Coulomb and HAL QCD strong interaction shown in the left panel of Fig.~\ref{fig:pXi} lies above the Coulomb predictions demonstrating the presence of an additional attractive strong interaction. These data provide a reference that can now be employed to test any theoretical calculation of the \pXim interaction.

The \pXim system presents two inelastic channels, n--$\Xi^0$ and $\Lambda$--$\Lambda$, just below threshold and three others, $\Lambda$--$\Sigma^0$, $\Sigma^0$--$\Sigma^0$ and $\Sigma^+$--$\Sigma^-$, well above threshold~\cite{Haidenbauer:2018jvl}. The latter, since their opening occurs far away from the \pXim mass threshold and theoretical predictions indicate a shallow interaction for these pairs, will have a negligible effect on the \pXim correlation function.
In the specific case of the  $\Lambda$--$\Lambda$ channel, precise femtoscopic measurements confirmed the weak strength of the strong interaction for these hyperons by means of hypernuclei data \cite{Takahashi:2001nm}.
Predictions based on chiral calculations for the n--$\Xi^0$ channel show a visible effect in the \pXim correlation signal~\cite{Haidenbauer:2018jvl} but currently this coupling is not yet present in lattice QCD calculations.
In the calculation carried out to obtain the theoretical correlation function shown by the pink histogram in the top panel of Fig.~\ref{fig:pXi}, the non diagonal terms of the interactions that contain the contribution of coupled-channels (see section \ref{subsec:coupledchannel}) are neglected.
Nevertheless, a direct measurement of these inelastic contributions is necessary to draw solid conclusions.

\subsection{Search for bound states}
\label{subsec:boundstates}

The $\Lambda$--$\Lambda$ interaction attracted the attention of both theoreticians and experimentalists already many years ago~\cite{Jaffe:1976yi} because of the possible existence of the H-dibaryon: a bound state composed of six quarks (uuddss). From an experimental point of view, the $\Lambda$--$\Lambda$ interaction was first addressed by studying the production of double-$\Lambda$
hypernuclei. The measurement of the BE of the hypernucleus $^6_{\Lambda\Lambda}$He allowed the estimation of the $\Lambda$--$\Lambda$ BE $=\, 6.91 \pm 0.6 $ MeV~\cite{Takahashi:2001nm}. This value was considered as an upper limit for the H-dibaryon. 
Also direct searches for the decays H$\rightarrow \Lambda \mathrm{p} \pi$ were carried out~\cite{Adam:2015nca}, but they never delivered any evidence.
A more recent upper limit evaluation of the bound state BE was obtained from a correlation analysis~\cite{FemtoLambdaLambda}.

It was also proposed to study such an interaction by analyzing heavy-ion collision data~\cite{Ohnishi:2016elb} and
the first attempt to investigate the $\Lambda$--$\Lambda$ final state via correlations was carried out by the STAR collaboration in Au\mbox{--}Au collisions at $\sqrt{s_\mathrm{NN}}=200$ GeV ~\cite{Adamczyk:2014vca}. This analysis delivered a scattering length and an effective range of $f_0^{-1}=\,-0.91\pm 0.31^{+0.07} _{-0.56} \,\, \mathrm{fm}^{-1}$ and $d_0=\,8.52\pm 2.56^{+2.09} _{-0.74} \,\, \mathrm{fm}$. These values correspond to a repulsive interaction. However, it was shown that the values and the sign of the scattering parameters strongly depend on the treatment of feed-down contributions from weak decays to the measured correlation. A re-analysis of the data outside the STAR collaboration extracted a positive value for $f_0^{-1}$ corresponding to a  shallow attractive interaction potential~\cite{Morita:2014kza}.
\begin{figure}[h!]
\centering
 \includegraphics[width=\textwidth]{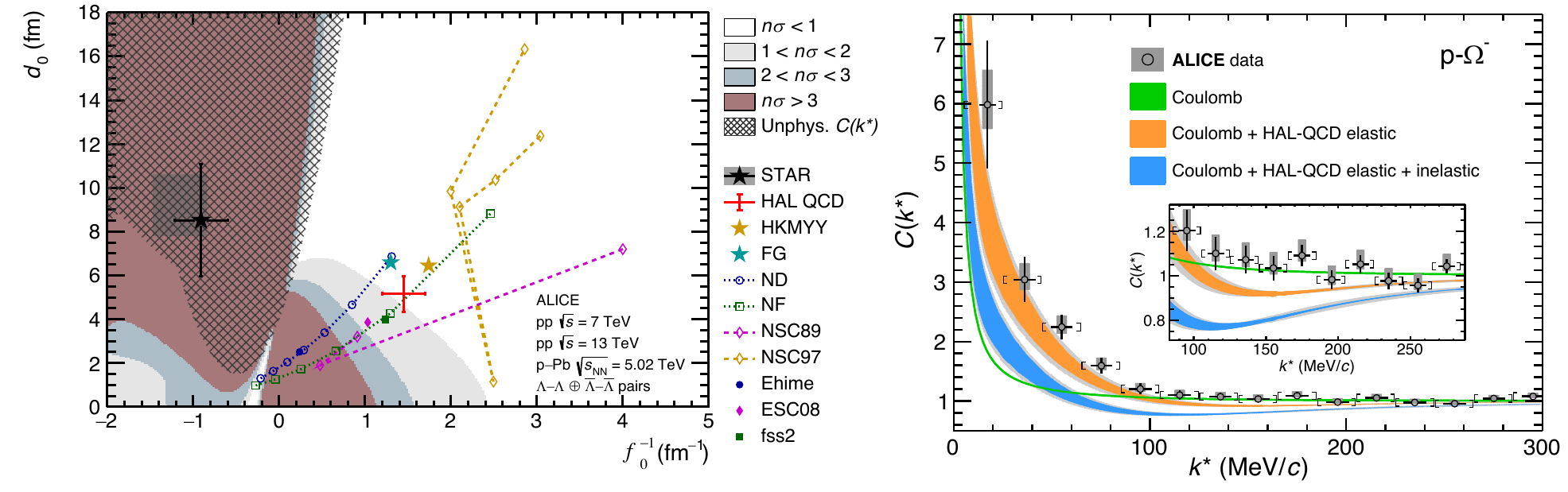} 
  \caption{\textit{(Color online)}
  Left panel: Exclusion plot of the scattering parameters for the $\Lambda$--$\Lambda$ interaction evaluated by testing the different values against the $\Lambda$--$\Lambda$ correlation.
  Right panel: Correlation function of  \pOm pairs measured by ALICE in high multiplicity \pp collisions at $\sqrt{s}=13$~TeV~\cite{Acharya:2020asf}.
  The data are shown by the black symbols, the systematic errors are shown by the grey boxes. The green line represents the expected correlation function by
  taking into account only the Coulomb interaction, its width is determined by the uncertainty in the source radius.
  The blue and orange bands consider both Coulomb and strong interaction by the HAL QCD collaboration~\cite{Iritani:2018sra}. The orange band considers for the strong interaction only the elastic contributions, the blue band considers elastic and inelastic contributions, its width represents the uncertainties associated with the lattice QCD calculations, and the grey band represents, in addition, the uncertainties associated with the determination of the source radius. The source radius, determined experimentally, is \radiuspOmega 
  fm.
  The inset shows in detail the correlation function around unity.}
  \label{fig:LL_pomega}
\end{figure} 
The $\Lambda$--$\Lambda$ correlations measured in pp and p-Pb collisions by ALICE at $\sqrt{s_\mathrm{NN}}= 7, 13$ TeV and $5.02$ TeV, respectively~\cite{FemtoRun1,FemtoLambdaLambda}
were also employed to study the interaction, and the residual correlations were treated by means of a novel data-driven method.
Since the statistics of the $\Lambda$--$\Lambda$ pairs with small relative momentum was limited, instead of extracting the scattering parameters from the fit of the correlation function a different approach was carried out ~\cite{FemtoLambdaLambda}.
A scan of different combinations of scattering parameters ($f_0^{-1}$ ,$d_0$) in the range $f_0^{-1} \in [-2,5]$ fm$^{-1}$ and $d_0 \in [0,18]$ fm was performed.
For each combination of values of the scattering parameters the correlation function was evaluated by using the Lednický-Lyuboshitz (LL) method.
The agreement with the experimental correlation function, using all data samples from pp collisions at $\sqrt{s}= 7, 13$ TeV and p-Pb collisions at $\sqrt{s_\mathrm{NN}}= 5.02$ TeV, is quantified in terms of a confidence level following the method in~\cite{NumericalRecipes}. 
The CATS framework is used to cross check the results from the LL method; the differences in the correlation functions obtained using the two methods are negligible. 
Also the gaussian source approximation, employed in the LL method, is validated by cross checks using the source profile predicted by the EPOS transport model~\cite{Pierog:2013ria,Mihaylov:2018rva} and considering the effects of short lived resonances.
The results, expressed in number of standard deviations ($n_{\sigma}$) are shown in the left panel of Fig. \ref{fig:LL_pomega}. 
The black hatched area represents the values for which the LL model breaks down for the small source sizes considered and delivers unphysical correlation functions. 
This region is unphysical because the resulting correlation function becomes negative.

This analysis has made it possible to extend the constraint to the scattering parameters and the BE of the  $\Lambda$--$\Lambda$ system.
The data is compared with models predicting either a strong attractive interaction~\cite{Ehime1}, a $\Lambda$--$\Lambda$ bound state~\cite{PhysRevD.15.2547,PhysRevD.20.1633}, or a shallow attractive interaction potential~\cite{ESC08,FG,HKMYY,Hatsuda:2018nes}.
Through the comparison shown in Fig.~\ref{fig:LL_pomega} one can see that the data favors a shallow attractive interaction, being compatible in particular with the~\cite{HKMYY} and~\cite{FG} models that are in agreement with hypernuclei data, and with the model~\cite{Hatsuda:2018nes}, that consists of preliminary lattice QCD calculations by the HAL QCD collaboration.
The data excludes the region corresponding to a strongly attractive or a very weakly binding short-range (small $|f_0^{-1}|$ and small $d_0$) interaction, and the first results from the STAR Collaboration~\cite{Adamczyk:2014vca} corresponding to a repulsive interaction are also excluded. 
The data does not exclude a $\Lambda$--$\Lambda$ bound state with a shallow binding (corresponding to negative $f_0^{-1}$ and small $d_0$ values). The upper limit of the BE for the H-dibaryon candidate can be evaluated applying the effective-range approximation and relating the parameters $f_0^{-1}$ and  $d_0$ to a corresponding BE via
\begin{align}\label{eq:ERE_BE}
 \mathrm{BE_{\Lambda\Lambda}} =\frac{1}{m_{\Lambda} d_0 ^2}\left ( 1-\sqrt{1+2 d_0f^{-1}_0}\right)^2.
\end{align}
Several caveats apply to this expression~\cite{FemtoLambdaLambda} since the implied effective range expansion might no be suited to described the bound state properties.
An analysis of the 1$\sigma$ region compatible with the existence of a bound state of the results shown in the left panel of Fig. \ref{fig:LL_pomega} was performed and allows a BE, considering statistical and systematic uncertainties, of BE = 3.2$^{+1.6}_{-2.4}$(stat)$^{+1.8}_{-1.0}$(syst) MeV~\cite{FemtoLambdaLambda}.

\vspace{2 mm}

An additional final state suited for the search for a baryon-baryon bound state is the \pOm channel.
Recent studies from phenomenological approaches~\cite{Sekihara:2018tsb} and first principle calculations~\cite{Iritani:2018sra} predict an attractive interaction potential at all distances between protons and $\Omega^-$ baryons. 
Both approaches also predict the existence of a \pOm bound state with binding energies of the order of a few MeV. For the strong-only and strong+Coulomb binding energies the two models predict $0.1$ MeV and $1$ MeV~\cite{Sekihara:2018tsb} and $1.54$ MeV and $2.46$ MeV~\cite{Iritani:2018sra}, respectively. 

As already pointed out in Section \ref{sec:Methodology}, the presence of a bound state manifests itself in a depletion of the correlation function with a strength that depends on the BE and the shape of the attractive potential.
In the presence of shallow bound states the correlation looks very similar as in the case of a strongly attractive interaction, but for more deeply bound state the correlation can also drop below the unity.

The \pOm correlation function has been recently studied by ALICE~\cite{Acharya:2020asf} using data from high-multiplicity \pp collisions at 13 TeV and the results are shown in Fig. \ref{fig:LL_pomega}.
The data, shown by the black points, is corrected for feed-down contributions and experimental effects, such as resolution effects at very small \ks values, meaning that the data can be directly compared to any theoretical prediction given a known emitting source.

The data in Fig. \ref{fig:LL_pomega} is compared with the predicted correlation function from calculations on the lattice by the HAL QCD collaboration~\cite{Iritani:2018sra} for a Gaussian source with a radius $r_0 =$ \radiuspOmega fm.
The source characteristics have been determined following the method explained in Sec.~\ref{eq:DecompSource_CoreRes}, for a $\langle m_{\mathrm{T}}\rangle$ of the \pOm pairs of 2.2 \GeVc, and taking into account the effect produced by short lived resonances.
The difference between the blue and orange colored bands corresponding to the HAL QCD prediction in Fig. \ref{fig:LL_pomega} reflect the current uncertainty of the calculations due to the presence of strangeness-rearrangement processes in the \pOm channel.

For the \pOm S-wave interaction, the total angular momentum $J$ can take on values of $J = 2$ or $J = 1$.
Processes such as $ \rm{p}\omm \rightarrow \Xi \La,\, \Xi\sig$ can occur~\cite{Morita:2019rph}, affecting the \pOm interaction in particular in the $J=1$ channel.
For the $J=2$ channel, the presence of strangeness-rearrangement processes should be strongly suppressed, since they are only possible through D-wave interaction processes. This $J=2$ channel is, so far, the only channel calculated by the HAL QCD Collaboration~\cite{Iritani:2018sra}. 
In order to compare the lattice QCD calculations with the ALICE data, two extreme assumptions are made for the description of the interaction in the $J = 1$ channel, following the recipe explained in \cite{Morita:2019rph}: i) no strangeness-rearrangement processes occur, and the shape of the $J = 1$ channel shows an attraction analogous to the $J=2$ channel as calculated by HAL QCD; ii) the
$J = 1$ channel is completely dominated by by strangeness-rearrangement processes, i.e. a complete absorption is assumed for this channel.
The correlation functions resulting from the assumptions i) and ii) are represented by the orange and blue lines in Fig. \ref{fig:LL_pomega}, respectively. For both predictions the Coulomb interaction is also taken into account, and the colored widths of the curves represent the intrinsic uncertainties of the lattice QCD calculations, with the grey curves showing, in addition, the uncertainties related to the experimental determination of the source radius. Clearly, the most attractive solution is preferred by the data, although the calculations underpredict the ALICE results at all \ks values.
In the absence of measurements of the $ \rm{p}\omm \rightarrow \Xi \La,\, \Xi\sig$ cross sections, future studies of $\La\mbox{--}\xim$ and $\Sigma^{0}\mbox{--}\xim$ correlations will help reducing the uncertainties in the expectations from the theory by pinning down the contributions of the inelastic channels. 

An evident depletion is present in the lattice QCD predictions 
shown in Fig. \ref{fig:LL_pomega}. By looking in particular to the region $k^*\in$~[100, 200]~\MeVc, one can see that the correlation function reaches values below the Coulomb-only prediction. Such depletion,
not confirmed by the experimental data,
is due to the presence of the \pOm di-baryon state.
The strength of the depletion depends on: i) the characteristics of the interaction; ii) the BE of the \pOm state; iii) the size of the particle emitting source. This dependence has been studied in detail in~\cite{Morita:2019rph} through the study of the interplay between the scattering length associated to the \pOm interaction and the corresponding correlation function obtained for different source sizes.

The ALICE data shown in Fig. \ref{fig:LL_pomega} do not follow the depletion predicted by the lattice QCD calculations. In order to obtain firm conclusions on the possible existence of the \pOm state, and, if existent, experimentally quantify its BE, a differential analysis of the \pOm correlations in systems with slightly different source sizes is necessary. This can be done at the LHC by ALICE by studying \pPb and peripheral Pb--Pb collisions.

\begin{figure}[h!]
 \includegraphics[width=0.999\textwidth]{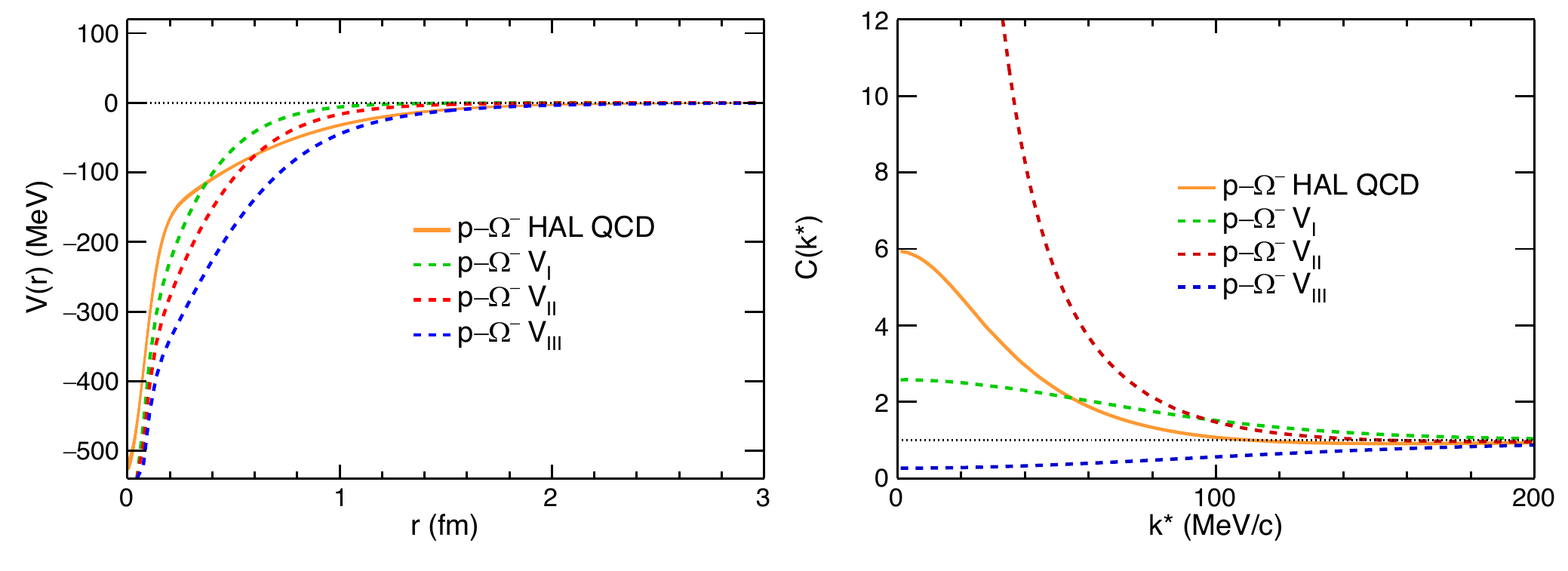}
  \caption{
  Left panel: Comparison of the strong interaction potentials for \pOm from references ~\cite{Morita:2016auo} (dashed lines) and ~\cite{Morita:2019rph} (orange solid line). The potentials , $V_{II}$ (red) and $V_{III}$ (blue) imply a \pOm bound state with binding energies due to strong interactions of  0.05 and 24.8 MeV. No bound state is associated with the $V_{I}$ (green) potential. The orange solid line represents the HAL QCD potential with nearly physical quark masses ~\cite{Morita:2019rph} predicting a BE of 1.54 MeV.
  Right panel: Correlation function for \pOm pairs corresponding to the potentials shown in the left panel for a radius of 0.95 fm.
  }
  \label{fig:viviiviii}
\end{figure} 

In order to show more clearly the effect of a possible bound state in the correlation function, it is useful to compare several local potentials describing the \pOm $J = 2$ interaction that are associated either to \pOm bound states with different properties or do not predict any bound state. We make use here of the potentials presented in~\cite{Morita:2016auo}, and labelled as $V_{I}$, $V_{II}$ and $V_{III}$. These potentials are based on dated calculations by the HAL QCD collaborations with non-physical quark masses~\cite{Etminan:2014tya} ($m_{\pi} = 875\,$ MeV, $m_{K} = 916$ MeV). For the construction of the $V_{II}$ potential, the lattice QCD data in~\cite{Etminan:2014tya} is fitted by an attractive Gaussian core plus and attractive Yukawa tail, while for the $V_{I}$ and $V_{III}$ potentials the range-parameter at long distance of the fit is varied in order to obtain a weaker and stronger attraction, respectively. The radial shape of such potentials compared with the most recent HAL QCD potential with physical quark masses~\cite{Morita:2019rph} ($m_{\pi} = 146 $ MeV, $m_{K} = 525$ MeV) can be seen in the left panel of Fig. \ref{fig:viviiviii}.
For $V_{I}$ no bound state is predicted, for $V_{II}$ a bound state with strong-only binding of 0.05 MeV  and strong+Coulomb of 0.63 MeV is predicted and for $V_{III}$ a bound state with strong-only binding of 24.8 MeV  and strong+Coulomb of 26.9 MeV is predicted \cite{Morita:2016auo}. 

The right panel of Fig. \ref{fig:viviiviii} displays the corresponding correlation functions for a source of $r_0 =$ \radiuspOmega fm. One can compare the limiting cases of a deeply bound \pOm bound state with a BE of 24.8 MeV ($V_{III}$ potential, blue curves) with the case of a very shallow BE of 0.05 MeV ($V_{II}$ potential, red curves) or no bound state ($V_{I}$ potential, green curves). Although the $V_{I}$ potential is much less attractive than the $V_{III}$ at all distances, the correlation function of the former is higher due to the deep depletion caused in the latter by the presence of a deeply bound state. Also the comparison between the $V_{II}$ and $V_{III}$ curves shows the same effect, since the shallow BE of the $V_{II}$ potential is reflected into a much shallower depletion as well. The correlation function corresponding to the most recent HAL QCD potential with physical quark masses~\cite{Morita:2019rph} is shown by the orange curve.
The BE of a few MeV is reflected in a correlation function that is lower than the one for the $V_{II}$ potential, despite the fact that the latter is less attractive for distances $r > 0.6$ fm.
In general a consistent picture is shown where for a very attractive interaction at all distances, the final correlation function for a small source size of around 1 fm, is determined to a high degree from the predicted BE of the bound state and the depletion caused by it.

The correlation function obtained from the $V_{I}$, $V_{II}$ and $V_{III}$ potentials for source sizes ranging from 2 to 5 $fm$ have been compared with data from ultrarelativistic \AuAu collisions at a center-of-mass energy of 200 GeV per nucleon pair by the STAR collaboration~\cite{STAR:2018uho}. 
The combination of a low purity and statistical significance of the data with 
such a large system size reduces the sensitivity of the comparison, as discussed at the end of section~\ref{sec:Methodology}.
The ratio of the correlation function for \pOm pairs in peripheral collisions (centralities of 40-80\%) to the one in central collisions (centralities of 0-40\%) at a $k^* = 20$ MeV/$c$ is compatible within 1$\sigma$ with the $V_{III}$, and within 3$\sigma$ with the $V_{I}$ and $V_{II}$ potentials for an expanding source.

\subsection{Coupled channel dynamics}
\label{subsec:coupledchannel}

Coupled-channel processes are widely present in hadron-hadron interactions whenever pairs of particles, relatively close in mass, share the same quantum numbers: baryonic charge B, electric charge Q and strangeness S. The coupling translates into on/off shell transitions from one system to the other.

Whenever present, the multi-channel dynamics deeply affect the hadron-hadron interaction and is at the origin of several phenomena, such as bound states and resonances, which crucially depend on the coupling between these inelastic channels.  A striking example can be found in the origin of the $\Lambda(1405)$, a molecular state arising from the coupling of antikaon-nucleon ($\bar{\mbox{K}}$--N) to $\Sigma$-- $\pi$~\cite{Hall:2014uca,Miyahara:2015bya}. In the baryon-baryon sector, the coupling between N--$\Lambda$ and N--$\Sigma$ is of great importance in providing the repulsive behaviour of \La hyperons in dense nuclear matter~\cite{Haidenbauer:2019boi}.

Since in femtoscopic measurements the final state is fixed (the measured particle pair), the corresponding correlation function represents an inclusive quantity able to show sensitivity to all the available initial inelastic channels produced in the collision~\cite{Haidenbauer:2018jvl,Kamiya:2019uiw}.

The effects of coupled-channels on the final measured correlation function depends on two main ingredients: the coupling constant strength, stemming from the strong multi-channel dynamics, and the conversion weights, namely the amount of pairs in the corresponding channel produced close enough to convert into the final measured state.
The correlation function in Eq.~\ref{eq:KooninPratt} needs to be modified and, for a system with $N$ coupled-channels, this observable in the $i$ channel that is measured reads~\cite{Lednicky:1981su,Haidenbauer:2018jvl,Kamiya:2019uiw} 

\begin{align}
\label{eq:corrfun_CC}
    C_i(k^*)
= \int d^3 r^* S_i (r^*)
|\psi_i (k^*_i,r^*)|^2 +
 \sum_{j\neq i}
 ^N w_j \int d^3 r^* S_j (r^*)
 |\psi_j (k_j^*,r^*)|^2.
\end{align}

Generally speaking, the emitting source $S_i (r^*)$ and the one for the incoming inelastic channels, $S_j (r^*)$, might be different (since the $m_{\mathrm{T}}$ distribution of the different pairs can differ) but the results presented in Sec.~\ref{sec:source} and the proximity in mass amongst the different channels quantitatively prove that the equality  $S_i(r^*)=S_j(r^*)$ can be assumed.

The first integral on the right-hand side describes the elastic contribution where initial and final state coincide, while the second integral is responsible for the remaining inelastic processes $j\rightarrow i$. The last integral depends on two main ingredients: the wave function $\psi_j (k_j^*,r^*)$ for channel $j$ going to the final state $i$ and the conversion weights $w_j$. The latter are directly related to the amount of pairs, for each inelastic channel, produced  in the initial collision which are kinematically available to convert into the final measured state. Estimates for these weights can be obtained using information on yields from statistical hadronization models~\cite{Vovchenko:2019pjl,Becattini:2001fg, Wheaton:2004qb} and on the kinematics of the produced pairs from transport models~\cite{Pierog:2013ria}.

As can be seen from Eq.~(\ref{eq:corrfun_CC}), the correlation function involves contributions from both elastic and inelastic components, and in principle, the scattering amplitude of the single-channel $i$ cannot be fully isolated from the inelastic contributions. Depending on the coupling strength, the coupled-channel contributions $j$ modify the \CF\ in two different ways, whether their opening (the minimum energy at which they can be produced) occurs below or above the production threshold of the considered pair (the reduced mass of the pair). Inelastic channels opening below threshold do not introduce any shape modification to the \CF, they just act as an effective attraction, increasing the signal strength of the correlation function. Channels appearing above threshold instead lead to a modification of the \ks\ dependence of the \CF\ in the vicinity of the opening, which is typically translated into a cusp structure, whose height is driven by the coupling strength.

\begin{figure}[h!]
\centering
\includegraphics[width=\textwidth]{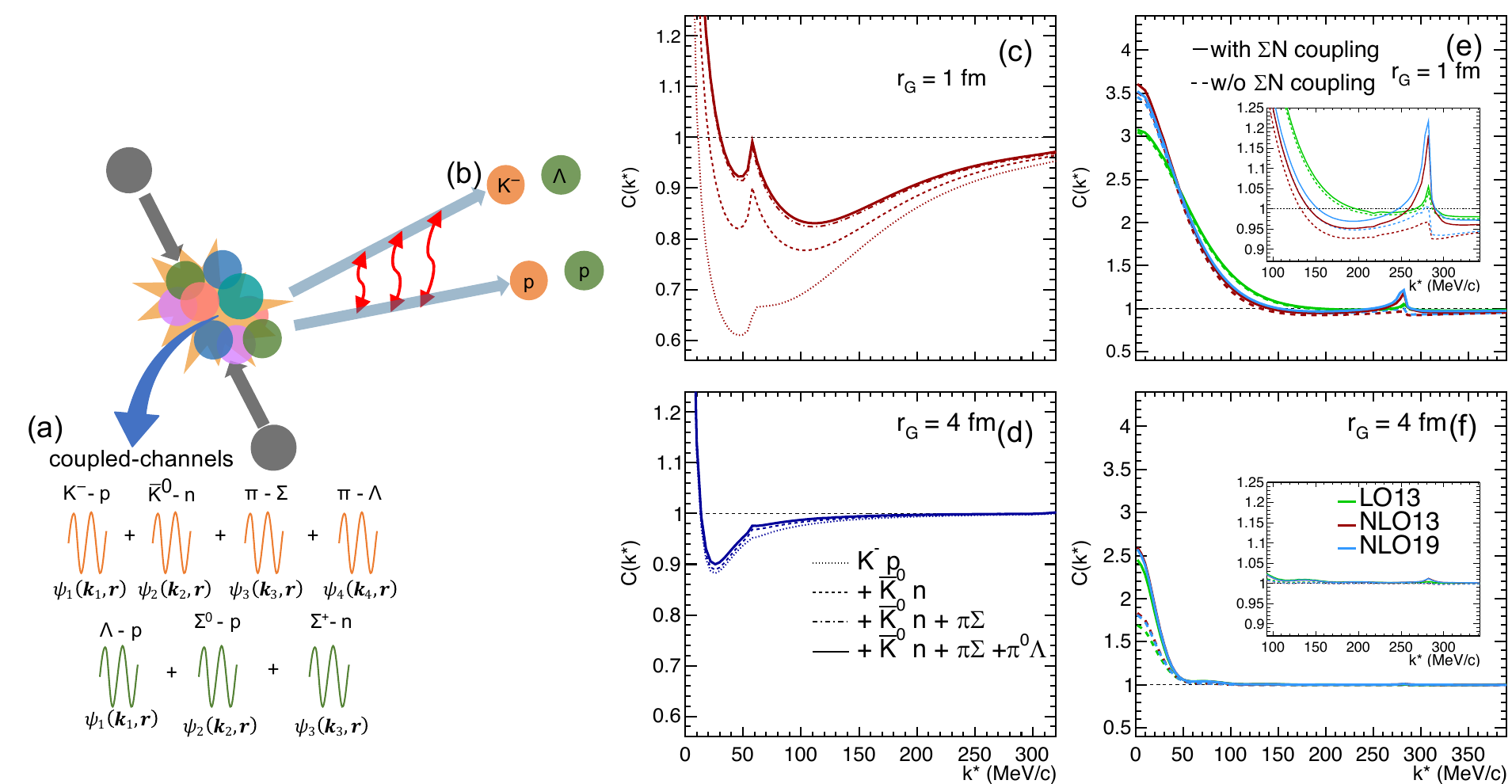}
\caption{\textbf{Schematic representation of the effects of
coupled-channels on the K$^-$--p and \pL  correlation function}. (a,b) System configuration in femtoscopic measurements, where only the final K$^-$--p and \pL channels are measured. Upper panels (c,e) : results for radii achieved in pp collisions (1 fm). Lower panels (d,f) : results for radii achieved in heavy-ion collisions (4 fm). In (c) and (d) the correlation function for K$^-$--p, from the pure elastic term (dotted line) to the full \CF (solid line) with all coupled-channels ($\bar{\mathrm{K}}^0$--n, $\pi$--$\Sigma$, $\pi$--$\Lambda$) included and the conversion weights are fixed to unity.  In (e) and (f) the \pL correlation function obtained assuming LO (LO13~\cite{Polinder:2006zh}) and two versions of NLO (NLO13~\cite{Polinder:2006zh},NLO19~\cite{Haidenbauer:2019boi}) \chiEFT calculations. Contributions from S-,P- and D-waves are included. Dashed lines: results with only the elastic term $\pL \rightarrow \pL$ in Eq.~\ref{eq:corrfun_CC}. Solid lines: results with the inclusion of coupled-channel contributions in Eq.~\ref{eq:corrfun_CC} from $\nSigp$ and $\pSigz$ with conversion weight $1/3$.}
\label{fig:FemtoBox_CC}
\end{figure}

These two main differences are illustrated in Fig.~\ref{fig:FemtoBox_CC} for the \Kmp and \pL systems.\\
The \Kmp system presents couplings to several inelastic channels below threshold such as $\pi\Lambda$, $\pi\Sigma$ and, due to the breaking of isospin symmetry, to charge-conjugated $\rm\bar{K}^0 N$ at roughly $4$ MeV above threshold corresponding to $k^*\approx 60$ \MeVc in the \CF.
On the left part (a) of Fig.~\ref{fig:FemtoBox_CC}, a schematic representation of the collision is shown. From the emitting source, formed after the collision, all the pairs constituting the four coupled-channels are produced and described by the corresponding wave functions $\psi_j(k_j^*,r^*)$.
The correlation of \Kmp pairs composing the final-state (channel 1) is measured (b) and its decomposition in the different channels contributions is shown in (c) and (d) for two different source sizes. The largest contributions to the \CF\ from coupled-channels occur for a small emitting source with $r_G=1$ fm in (c).
The \CF\ signal increases as the inelastic contributions are added and the cusp structure, visible when the \Kzn channel is explicitly added, indicates the opening of this channel above threshold. 
For both source radii, this structure already appears when the mass difference between \kam and $\rm \bar K ^0$ is considered, and it is present also in the elastic $\rm K^-p$$\rightarrow$$\rm K^-$p contribution (dotted line).
 As already mentioned, the explicit inclusion of the $\rm \bar{K}^0$--N contribution acts as an "effective" attraction component, increasing the signal of the correlation function and of the cusp accordingly to the strength of the coupling between the two channels (short-dashed line).
This effect is suppressed when the source size is increased up to $r_G=4$ fm (d), as in central heavy-ion collisions.

A similar trend can be seen when another strongly coupled-channel is introduced, the $\pi$--$\Sigma$ (dashed-dotted line), responsible for the dynamic generation of the molecular state $\Lambda (1405)$.
The strong coupling to this channel, lying below threshold, is directly translated into the correlation function of K$^-$--p pairs, visible as a clear enhancement of the signal at low momentum with respect to the single-channel contribution.\\
The extreme sensitivity to coupled-channel contributions of the \CF\ obtained in small systems has been recently confirmed by results for the K$^-$--p correlation function, measured by the ALICE Collaboration in pp collisions at different energies~\cite{Acharya:2019bsa}. Future measurements of this pair, performed in different colliding systems, will also be able to provide quantitative constraints on the coupling strength to the $\rm \bar K ^0$n channel.\\
In Fig.~\ref{fig:FemtoBox_CC} another coupled system, formed by the interaction of respectively a \La and a $\Sigma$ with nucleons, is depicted. The strength in the \NLa$\leftrightarrow$\NSig conversion is not experimentally well constrained since scattering measurements cannot currently provide precise enough data on the \pL cross section at momenta close to the opening~\cite{Eisele:1971mk,Alexander:1969cx,SechiZorn:1969hk}.  The only experimental observations of the $\Sigma$--p cusp have been extracted in partial-wave analyses on p+p$\rightarrow$ pK$^+ \Lambda$ reactions at low energy but they are strongly affected by $\Lambda$--p final-state interactions~\cite{ElSamad:2012kg,Munzer:2017hbl}.\\
In Fig.~\ref{fig:FemtoBox_CC} the theoretical \pL correlation function obtained within different calculations based on \chiEFT, next-to-leading (NLO)~\cite{Polinder:2006zh,Haidenbauer:2019boi} and leading order (LO)~\cite{Polinder:2006zh}, is shown for a source radius of 1 fm and for a momentum cutoff parameter of 600 MeV/$c$.
The coupling to the \NSig (\nSigp, \pSigz) occurs already in the S wave and finds the largest contribution from D waves, hence partial waves up to $l=2$ are included. The conversion weights $w_j$ for this coupling in Eq.~\ref{eq:corrfun_CC} can be fixed to $1/3$ from isospin symmetry, a value comparable with thermal model calculations~\cite{Vovchenko:2019pjl} and measurements of production ratios between these two hadrons at high energies~\cite{Borissov:2019dcv,VanBuren:2005sk}.
The inclusion of the \NSig coupled-channel contributions, as shown for the K$^-$--p case, leads to the appearance of a cusp structure at $\ks=289$ \MeVc, corresponding to the kinematic opening of the inelastic \nSigp and \pSigz channels.
The low momentum region of the \CF\ is not deeply affected by the explicit inclusion of the inelastic terms since the opening of the \NSig occurs above threshold.\\
The largest differences in the behaviour of the \CF, regardless of the presence or absence of \NSig contributions, arise in the LO and NLO descriptions.
The LO predictions have already been ruled out by scattering data in the proximity of the cusp region, since the calculation deviates significantly from the data, despite the large uncertainties. 
The two versions of the NLO calculations (NLO13, NLO19) mainly differ in the description and strength of the \La$\leftrightarrow\Sigma$ conversion potential, which leads to significant modifications of the \La hyperon interaction in dense nuclear matter and to different results for light hypernuclei. \\
As can be seen in Fig.~\ref{fig:FemtoBox_CC}, the cusp height predicted from these two approaches is rather similar but mild differences are present below and above the cusp. 
The high precision data delivered by the recent ALICE measurements on \pL pairs~\cite{ALICEpLambda} favoured the latest NLO19 chiral potential, indicating a weak coupling between \NLa and \NSig channels and predicting a more attractive \La single-particle potential in neutron matter.
This current picture on the \pL interaction have profound implications for three-body hyperonic forces. This scenario is also directly relevant to open problems in astrophysics such as the presence of hyperons in NSs~\cite{Lonardoni:2014bwa,Logoteta:2019utx,Gerstung:2020ktv}.\\
In conclusion, femtoscopic measurements in pp collisions are able to probe the short--distance region of the pair wave-function, in which the coupled-channel dynamics dominate the strong interaction. Moving to larger source sizes tests the asymptotic part of the wave-function where the inelastic terms are noticeably suppressed and partial access to the pure elastic interaction can be obtained. 
This opens the possibility to investigate the dynamics of the couplings between the elastic and inelastic channels by performing femtoscopic measurements of the same pair in different colliding systems, leading to a complete description of all hadron-hadron interactions within SU(3).\\

\section{Implications for neutron stars}
\label{sec:neutronstars}
The interaction of hyperons with nucleons is one of the key ingredients needed to understand the composition of the most dense objects in our universe: neutron stars (NS)~\cite{Ozel:2016oaf,Riley:2019yda}.
These kinds of stars are the final outcome of supernova explosions and are typically characterized by large masses ($M\approx 1.2-2.2M_\odot$) and small radii ($R\approx 9-13$ km)~\cite{Demorest:2010bx,Antoniadis:2013pzd,Cromartie:2019kug}. 
In the standard scenario, the gravitational pressure is typically counter-balanced by the Fermi pressure of neutrons in the core, which, along with electrons, are the only remnants from the mother-star collapse. The high density environment ($\rho\approx3-4\rho_0$) supposed to occur in the interior of NS leads to an increase in the Fermi energy of the nucleons, translating into the appearance of new degrees of freedom such as hyperons. This energetically favored production of strange hadrons induces a softening of the Equation of State (EoS).\begin{marginnote}
\entry{Equation of State}{Thermodynamical relation between  pressure, energy-density and temperature describing the properties of nuclear matter under extreme conditions (high T or high density) and strongly dependent on the constituents and the interactions among them.}
\end{marginnote}The behavior of the mass as a function of the radius has a unique correspondence with the EoS through the solution of the Tolman–Oppenheimer–Volkoff equations, hence the mass-radius relation strongly depends on the constituents of the EoS and on their interactions.
The inclusion of hyperons leads to NS configurations unable to reach the current highest mass limit from experimental observations of $2.2M_\odot$~\cite{Cromartie:2019kug}.\\
For this reason, the presence of hyperons inside the inner cores of NS is still under debate, and this so-called hyperon puzzle is far from being solved~\cite{Djapo:2008au,Tolos:2020aln}.\\
A key element in the complete understanding of this puzzle is the interaction of hyperons with the surrounding medium, which strongly affects the properties of the corresponding EoS~\cite{Weissenborn:2011kb,Weissenborn:2011ut} and can be related to the interaction between hyperons and nucleons (Y--N and Y--N--N) in vacuum.
A repulsive Y--N interaction occurring already at the two-body level can push the appearance of hyperons to larger densities, limiting the possible presence of these particle species inside NS, stiffening the EoS and allowing for larger star masses.\\
The more precisely the hyperon-nucleon two-body and three body interactions are known in vacuum, the more detailed is the knowledge of the hyperonic content inside NS.
A large interest in this topic has been triggered by the recent measurements of gravitational waves signals from NS mergers, which opened a new gate to experimentally access the properties of the matter inside NS.\\
As already shown in the previous sections, femtoscopy is capable of providing new insight into interactions involving nucleons and hyperons which are poorly understood or not accessible with scattering experiments.\\
A key example is given by the femtoscopic measurements of the \pL strong interaction.
The $\Lambda$ baryons are typically the first hyperon species that are produced inside NS, due to their light mass. Their appearance is also theoretically favored by the overall attractive potential that a $\Lambda$ feels at the saturation density, $U_\Lambda = -30$ MeV~\cite{Hashimoto:2006aw}. The results obtained on this system, as discussed in Sec.~\ref{subsec:coupledchannel}, support recent \chiEFT calculations in which an even more attractive interaction of the \La with the surrounding nucleons, due to the \LN$\leftrightarrow$\SN dynamics, is predicted. In this case the early appearance of \La hyperons in neutron matter will lead to a too soft EoS and ultimately to stable light NS configurations. Such a scenario, in order to co-exist with the astrophysical constraints on NS masses, requires the introduction of repulsive forces that might be  present in other YN systems and in the inclusion of three-body interactions.\\
Repulsive Hyperon-nucleon-nucleon interactions, such as $\Lambda$--N--N, have already been included in several approaches to obtain a stiffer EoS~\cite{Haidenbauer:2016vfq,Lonardoni:2014bwa}.
However, at the moment, these three-body forces rely on the experimental measurements of hypernuclei binding energies ($^4 _\Lambda \mathrm{H}$, $^4 _\Lambda \mathrm{He}$), in which the determination of the genuine $\Lambda$--N--N interaction is not straightforward and can be affected by many-body effects.
For this reason the current theoretical understanding of the role played by three-body terms in the strangeness $|S|=1$ sector inside NS is not yet settled.\\
A major improvement in the understanding of the role played by heavier strange hadrons in the hyperon puzzle has been achieved by the validation of lattice QCD predictions for the \NXi interaction.
\begin{figure}[t!]
\centering
\includegraphics[width=\textwidth]{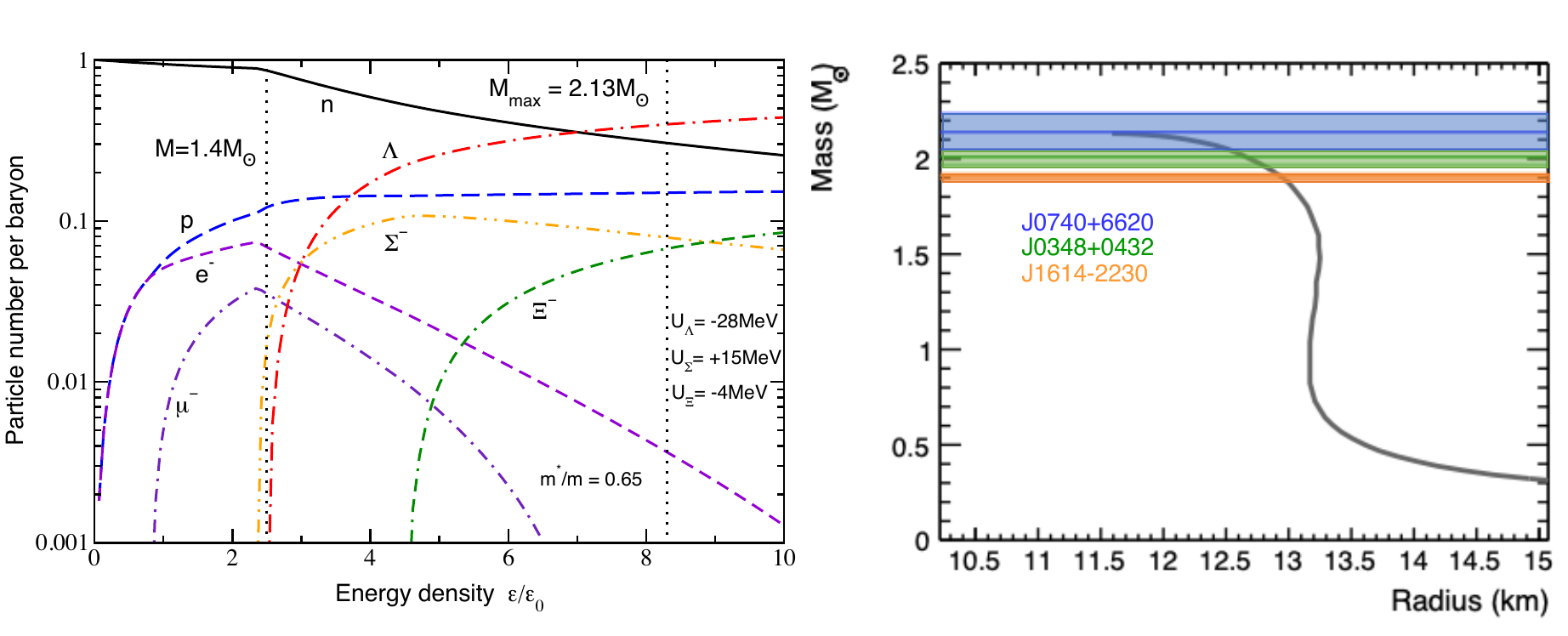}
\caption{Left: fraction of particles produced in the inner core of a NS as a function of the energy density, in units of energy density $\epsilon_0$ at the nuclear saturation point. The single-particle potential depths in symmetric nuclear matter (SNM) for $\Lambda$, $\Sigma$ and $\Xi$ hyperons are displayed.  The vertical dotted lines indicate the central energy densities reached for a standard NS of $1.4M_\odot$ and for the maximum mass, $2.13M_\odot$, reached within this specific EoS. The mean-field calculations~\cite{Schaffner:1995th,Weissenborn:2011kb,Weissenborn:2011ut,Hornick:2018kfi} have been tuned in order to reproduce the lattice predicted value of $U_\Xi$ in pure neutron matter (PNM) obtained in~\cite{Inoue:2016qxt}, using the in-vacuum results validated by ALICE data in ~\cite{FemtopXi}. The EoS obtained with these constraints provides a stable NS with a maximum mass of $M_{\rm max} = 2.13M_\odot$, as seen on the mass-radius plot on the right and is compatible with recent astrophysical measurements of heavy NS, indicated by the blue~\cite{Cromartie:2019kug}, green~\cite{Antoniadis:2013pzd} and orange~\cite{Demorest:2010bx} bands. }
\label{fig:NS_fractions}
\end{figure}
As shown in Sec.~\ref{subsec:hypnuclint}, the measurement of the \pXim correlaion~\cite{FemtopXi} confirmed a strong attractive interaction between these two hadrons and provided a direct confirmation of lattice potentials~\cite{Sasaki:2019qnh}.
Using this same interaction as a starting point to extrapolate results in a neutron-rich dense system, a repulsive average interaction of roughly $+6$ MeV can be obtained~\cite{Inoue:2016qxt}. Currently, models for EoS including $\Xi$ hyperons assume large variations in the values of the single particle potential ($-40,+40$ MeV)~\cite{Weissenborn:2011kb} and hence the
validated lattice predictions impose a much more stringent
constraint.
In Fig.~\ref{fig:NS_fractions}, the fractions of particles, obtained from mean-field
calculations~\cite{Schaffner:1995th,Weissenborn:2011kb,Weissenborn:2011ut,Hornick:2018kfi},
produced in the inner part of NS are shown as a function of the energy density.
The single-particle potentials for $\Lambda$ and $\Sigma$ hyperons have been fixed to the current values constrained from scattering data and hypernuclei, and confirmed by the LHC measurements. 

The isovector couplings to the $\Xi$ have been adjusted to reproduce the predicted results in pure neutron matter (PNM) obtained from HAL QCD calculations at finite density~\cite{Inoue:2016qxt}, stemming from the predictions in vacuum discussed in Sec.~\ref{subsec:hypnuclint}.
The slight repulsion acquired by a $\Xi^-$ in pure neutron matter directly translates into larger energy densities, and hence larger nuclear densities, for the
appearance of this hyperon species. In the right plot of Fig.~\ref{fig:NS_fractions}, the resulting mass-radius relation obtained by assuming the predicted HAL QCD $\Xi$ interaction in medium is shown. The production of cascade hyperons occurring at higher densities leads to a maximum NS mass of $2.13M_\odot$, fully compatible with the recent measurements, indicated by the colored bands, of NS close to and above two solar masses~\cite{Demorest:2010bx,Antoniadis:2013pzd,Cromartie:2019kug}.\\

Recent results in small colliding systems have proven that femtoscopy can play a central role in understanding the dynamics amongst hyperons and nucleons in vacuum. Comparisons between hadronic models and these data are necessary in order to constrain calculations at finite density and to pin down the hyperons behaviour in a dense matter environment. 
The unique possibility to investigate different YN interactions and to extend the measurements to three-body forces, can finally provide  quantitative input to the long-standing hyperon puzzle.

\section{Outlook}
\label{sec:outlook}

A complete program of new measurements in \pp collisions at 14 TeV has been approved for the upcoming Runs 3 and 4 of the LHC with ALICE~\cite{ALICE-PUBLIC-2020-005}. To address further questions regarding two- and three-body forces that involve hyperons, correlation studies constitute one of the main foci of such program.
Studies will benefit from data-taking with increased instantaneous luminosity and readout speed, plus better tracking and vertexing performances of the upgraded apparatus.
Moreover, the new data acquisition system will make it possible to select events with very high multiplicity, up to 16 times the average multiplicity of minimum bias pp collisions. Accessing such a regime of multiplicities in \pp events is particularly beneficial for measurements including strange hadrons due to the enhanced production of strangeness in collisions with high-multiplicity~\cite{ALICE:2017jyt}.
Assuming an acquired luminosity of 200 pb$^{-1}$ and a selection of events with a number of produced charged particles (\nch) seven times higher than the mean number of charged particles in minimum-bias collisions \nch $>$ 7 \meannch, an overall increase up to a factor 50 for particle pairs per event is expected for the Run 3 high-multiplicity data with respect to the sample collected in Run 2~\cite{ALICE-PUBLIC-2020-005}. 

Several new analyses can be performed with the Run 3 and Run 4 data that were not possible with the Run 1 and Run 2 statistics, and the question of three-body forces including hyperons can finally be experimentally addressed.

\addtocontents{toc}{\protect\setcounter{tocdepth}{1}}

\subsection{K$^{-}$--d correlations}
Following the measurement of K$^{-}$--p correlations in \pp ~\cite{Acharya:2019bsa}, the study of the correlation function of K$^{-}$--deuteron pairs will be realized. Together with the planned measurements at threshold using kaonic atoms by SIDDHARTA-2~\cite{Curceanu:2013bxa}, the K$^{-}$--d femtoscopy will allow us to determine for the first time the full isospin dependence of the $\mathrm{\overline{K}}$--N interaction, a fundamental problem in the strangeness sector in the low-energy regime of QCD.
    
\subsection{p\mbox{--}$\Sigma^0$ correlations}
The investigation on the p\mbox{--}$\Sigma^0$ correlation will provide precise data on an interaction that, in contrast to the N--$\Lambda$ interaction, is currently very poorly known experimentally. A first measurement~\cite{Acharya:2019kqn} of this correlation was performed, using high multiplicity \pp collisions, demonstrating the feasibility of the approach, although with large statistical uncertainties and relatively low signal purity. The minimum-bias \pp \forteen Run 3 data will allow to have a yield of p\mbox{--}$\Sigma^0$ pairs ten times higher than in the Run 2 data, that will deliver the first precise data in the field. At the same time lattice QCD calculations are expected to reach precision also in the $|S| = 1$ sector within the next few years, and hence the new measurement can contribute to the validation of the state-of-the-art theoretical calculations.

\subsection{$\Lambda$--$\Xi$ correlations}
The enhancement of the yield of strange particles in the high-multiplicity data from Run 3 will allow us to study as well the $\Lambda$--$\Xi$ interaction with high precision. Such a study will complement the measurements of \pOm correlations, and the comparison to the lattice QCD calculations. For the \pOm interaction, the $J=1$ channel lacks of any prediction so far, since it is dominated by absorption in the $\Lambda$--$\Xi$ and $\Sigma$--$\Xi$ channels.  The study of the $\Lambda$--$\Xi$ will hence provide the first experimental constraints to the contribution of the coupled channels for the \pOm system.

\subsection{$\Omega$--$\Omega$ correlations}
The HAL QCD collaboration has provided Lattice QCD calculations at the physical point for the $\Omega$--$\Omega$ system, predicting the existence of ''the most strange dibaryon'' with a BE of around 1.6 MeV~\cite{Gongyo:2017fjb}, implied by the strong attractive  character of the $\Omega$--$\Omega$ strong interaction and the fact that the Pauli principle does not apply for this system. 

So far, no experimental data is available for this interaction. The measurement of the $\Omega$--$\Omega$ correlation function is extremely challenging and constantly requested by theoreticians. During Run 3, the data acquisition of ALICE will be implemented with a dedicated trigger for $\Omega$ decays that can sample the whole $200$\,pb$^{-1}$ of the \pp data taking, resulting in a total of $2\cdot10^{9}$ reconstructed and recorded $\Omega^- \oplus \Omega^+$. For the correlation studies, a total of about $500$ $\Omega$--$\Omega$ pairs are expected to be reconstructed with low relative momentum, $k^* < 200$\,\MeVc.

On the theoretical side, the predictions from HAL QCD Collaboration~\cite{Morita:2019rph} provide the $^{1}S_{0}$ channel of the $\Omega$--$\Omega$ interaction alone.
This is the channel with the smallest contribution (weight 1/16), and an attractive interaction is also expected for the $^{5}S_{2}$ channel (weight 5/16), but the calculations are not available yet.

\begin{figure}[htb]
\centering 
 \includegraphics[width=0.58\textwidth]{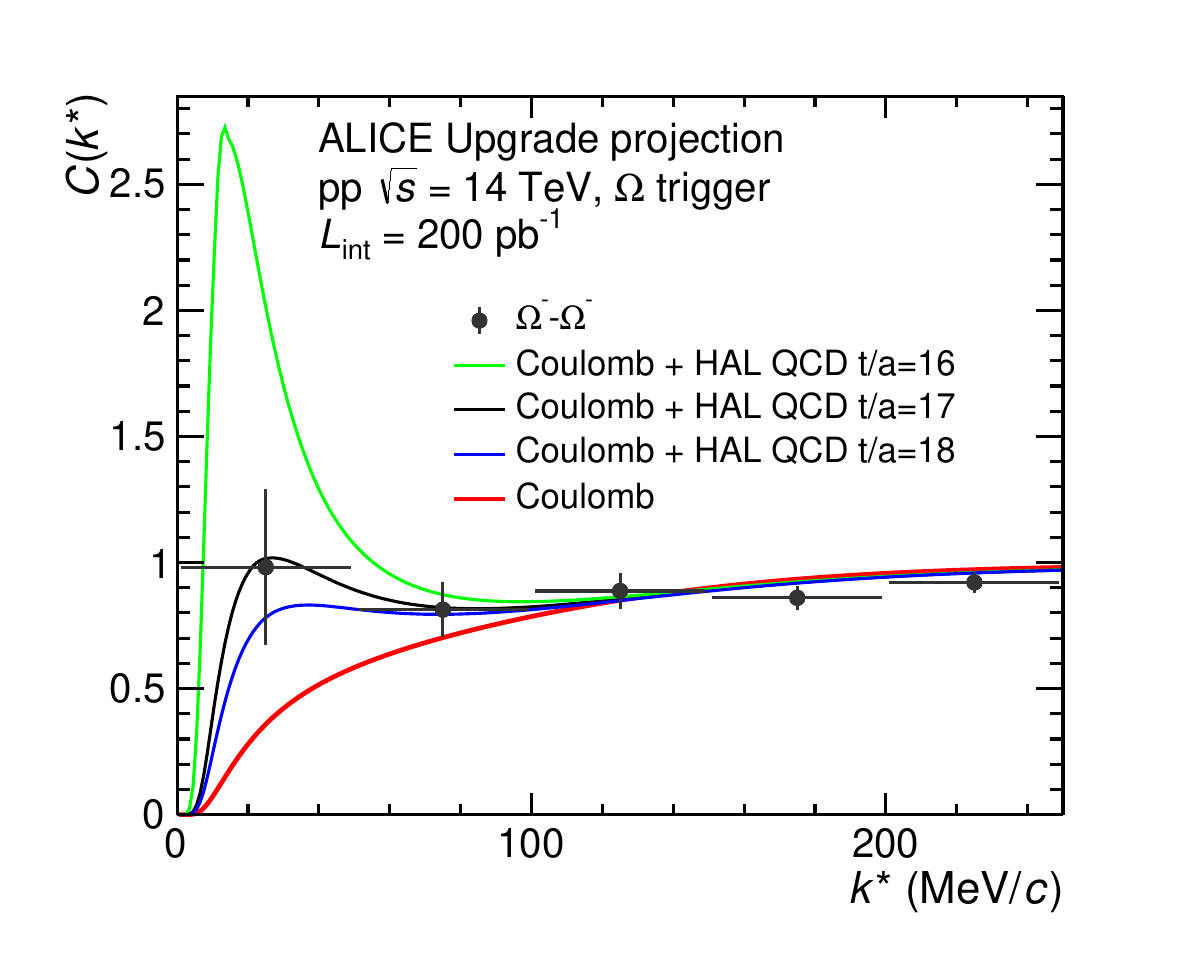}
  \caption{Expected precision of the $\Omega$-$\Omega$ correlation function with ALICE Run 3 data (black points).
   The red curve represents the Coulomb-only scenario, and the green, black and blue lines contain, in addition, the strong interaction potential by the HAL QCD collaboration with different integration times, t/a = 16,  t/a = 17,  and t/a = 18, respectively.
   The gaussian source used for the calculations has a radius of $ r = 0.8 $fm.
   The simulated data follows the Coulomb +  HAL QCD t/a= 17 scenario.}
  \label{fig:OmegaOmega}
\end{figure}

A projection of the measurement of the $\Omega$--$\Omega$ correlation is represented by the black points in Fig. \ref{fig:OmegaOmega} (left panel), and it is compared with the Coulomb-only scenario (red curve), and three different curves with the additional strong interaction predicted by the HAL QCD Collaboration. 
One can see that for the lattice calculations ($^{1}S_{0}$ channel) there are substantial differences in the expected correlation function considering different integration times (t/a parameter).
This is true in particular for a very small source like the one expected for $\Omega$--$\Omega$ pairs in pp collisions, with a radius of around $ r = 0.8 $ fm, 
and it is shown by the green, black and blue lines, that represent the calculations for t/a = 16,  t/a = 17, and t/a = 18, respectively.
The simulated data follows the Coulomb +  HAL QCD t/a= 17 scenario. 

 A precision of 31\% at a $k^*=25$\,\MeVc is expected in the correlation function using a bin width of 50 \MeVc. This would constitute a groundbreaking measurement of the $\Omega$--$\Omega$ interaction, delivering the first constraint on the lattice QCD calculations and with the potential for experimentally determining for the first time ever the sign of the strong interaction between $\Omega$--$\Omega$ pairs.

\subsection{$\Lambda$--d correlations}
Complementing the studies on $\Lambda$--p correlations, the study of the $\Lambda$--d correlation function provides additional information on the $\Lambda$--N interaction.
Experimental access to information on direct $\Lambda$--d scattering is even harder to obtain than that from $\Lambda$--p scattering, and correlation studies will constitute an additional and independent source of information for this channel. Moreover the measurement of the $\Lambda$--d correlation function in small systems complements the measurement of the hypertriton BE and delivers information on many-body forces~\cite{Haidenbauer:2020uew}.
    
There are two different spin configurations in the S-wave $\Lambda$--d interaction: the doublet $^2S_{1/2}$ and the quartet $^4S_{3/2}$ states. With no scattering data available for the $\Lambda$--d channel, the scattering parameters in the doublet state are constrained by measurements of the lifetime of the bound state found in this partial wave, the hypertriton, $^3$H$_\Lambda$. The hypertriton BE is related to the scattering parameters in the effective range approximation via the Bethe formula~\cite{Bethe:1949yr}. The higher spin configurations are not binding, and they are currently not tested by any experimental data. For the constraint of the quartet state, chiral SU(3) calculations~\cite{Haidenbauer:2013oca} are used so far.
    
The expected precision of the measurement of the  $\Lambda$--d correlation function with the \pp \forteen high multiplicity data sample with \nch $>$ 7 \meannch during Run 3 at $k^* = 50$\,\MeVc is on the order of 5\% (with a bin width of 20 \MeVc). Such a precise measurement will complement the hypertriton BE measurements, scanning the full spin dependence of the $\Lambda$--N interaction and, as has been suggested, possibly providing as well an insight into the coalescence process. These studies in pp can be complemented by studies in larger systems (Pb--Pb) leading to a better knowledge and understanding of many-body forces acting in light hypernuclei~\cite{Haidenbauer:2020uew}.

\subsection{Three-body forces}
In addition to the $\Lambda$--deuteron measurement, the study of three-body interactions involving hyperons, that are extremely important for the understanding of the structure of neutron stars, will be experimentally accessible with high precision for the first time.
Exclusive measurements of p--p--$\Lambda$ and correlations with a newly developed mathematical formalism using cumulants that allow the study of the correlation function of three particles with non-identical masses will enable accessing the final state interaction. For this purpose, the data acquisition will be implemented to sample the whole data taking recording events where at least two proton candidates and one $\Lambda$ candidate are reconstructed using an on-line trigger selection.
 
\addtocontents{toc}{\protect\setcounter{tocdepth}{2}}

\section{Summary}
\label{sec:summary}
 The correlation technique employed to study the strong interaction among hadrons has been discussed. Since a precise understanding of the source that characterizes the particle emission is mandatory to extract the strong interaction from correlations, a dedicated model for small colliding systems at the LHC has been motivated and explained. In this model, the contribution from the strong decay of short-live resonances has been modelled and the hypothesis of an universal source for all hadron-hadron pairs has been demonstrated and exploited in all the discussed analyses. The method has been tested employing p--p and p--$\Lambda$ correlations, where the interaction is rather well known, especially for p--p pairs. Following this scheme the results achieved for several hyperon--nucleon, hyperon--hyperon and kaon--nucleon combinations have been presented.
The first measurement of the attractive p--$\Xi^-$ strong interaction was presented and confirmed by lattice calculations by the HAL QCD Collaboration. The precise measurement of the $\Lambda$--$\Lambda$ interaction has allowed for extraction of the most precise upper limit for the BE of a possible H dibaryon state. The first measurement of the attractive strong p--$\Omega^-$ interaction has been shown as well. In the latter case, the results compared with lattice calculations for the first time do not show yet any clear evidence for the existence of a bound state. It has been shown that the presence of the coupled channel dynamics in the K$^-$--p and p--$\Lambda$ channels  manifests itself on the correlation functions. The coupling $\rm \bar{K}^0$--N$\leftrightarrow$K$^-$--p and \NLa$\leftrightarrow$\NSig have been directly observed for the first time. 

The consequences of the new measurement of the strong interaction among protons and strange hadrons for the physics of neutron stars has also been addressed. The example related to the $\Xi^-$ strong interaction has shown the impact of the new measurements on astrophysics.
In the future, the measurements of additional two-body correlations and possibly  three-body correlations among hyperons and nucleons are planned and, if achieved, they will impose more stringent constraints for neutron stars.
In particular, the physics opportunities that will be available during the LHC Run 3 data taking have been sketched. \\
In general, it was demonstrated that the correlation technique applied to small colliding systems at the LHC is a very promising tool with which to investigate the strong interactions. For this reason, a new laboratory to study hadron-hadron interaction has been established with the capability to unveil the strong interaction among any hadron--hadron pair.

\section*{DISCLOSURE STATEMENT}
The authors are not aware of any affiliations, memberships, funding, or financial holdings that
might be perceived as affecting the objectivity of this review. 

\section*{ACKNOWLEDGMENTS}
The authors would like to thank Dimitar Mihaylov, Alice Ohlson and Thomas Humanic for their inputs and comments to this manuscript. 
A special thanks also to Juergen Schaffner-Bielich, Debarati Chatterjee, Suprovo Gosh and Benjamin Doenigus for the interesting discussions on neutron stars physics.
This work has been supported by DFG EXC 2094 390783311 ORIGINS, GSI TMLRG1316F, BmBF 05P15WOFCA, SFB 1258, DFG FAB898/2-2.

%

\bibliographystyle{ar-style5.bst} 
\bibliography{Bibliography.bib}

\end{document}